\documentclass[11pt]{article}
\usepackage{amsfonts,amsmath,amsxtra}
\usepackage{amssymb,amscd}
\usepackage[all]{xy}
\catcode`\@=11
\def\marginnote#1{}
\newcount\hour
\newcount\minute
\newtoks\amorpm
\hour=\time\divide\hour by60
\minute=\time{\multiply\hour by60 \global\advance\minute by-\hour}
\edef\standardtime{{\ifnum\hour<12 \global\amorpm={am}%
     \else\global\amorpm={pm}\advance\hour by-12 \fi
     \ifnum\hour=0 \hour=12 \fi
   \number\hour:\ifnum\minute<10 0\fi\number\minute\the\amorpm}}
\edef\militarytime{\number\hour:\ifnum\minute<10 0\fi\number\minute}
\def\draftlabel#1{{\@bsphack\if@filesw {\let\thepage\relax
\xdef\@gtempa{\write\@auxout{\string
   \newlabel{#1}{{\@currentlabel}{\thepage}}}}}\@gtempa
\if@nobreak \ifvmode\nobreak\fi\fi\fi\@esphack}
     \gdef\@eqnlabel{#1}}
\def\@eqnlabel{}
\def\@vacuum{}
\def\draftmarginnote#1{\marginpar{\raggedright\scriptsize\tt#1}}
\def\draft{\oddsidemargin -0.1truein
     \def\@oddfoot{\sl preliminary draft \hfil
     \rm\thepage\hfil\sl\today\quad\militarytime}
     \let\@evenfoot\@oddfoot \overfullrule 3pt
     \let\label=\draftlabel
     \let\marginnote=\draftmarginnote
\def\@eqnnum{{\rm (\theequation)}
\rlap{\kern\marginparsep\tt\@eqnlabel}%
\global\let\@eqnlabel\@vacuum}  }
\def\numberbysection{\@addtoreset{equation}{section}
     \def\theequation{\thesection.\arabic{equation}}}
\numberbysection

\renewcommand{\theequation}{\thesection.\arabic{equation}}
\makeatletter
\newdimen\normalarrayskip            
\newdimen\minarrayskip               
\normalarrayskip\baselineskip
\minarrayskip\jot
\newif\ifold             \oldtrue            \def\new{\oldfalse}
\def\arraymode{\ifold\relax\else\displaystyle\fi}
\def\eqnumphantom{\phantom{(\theequation)}} 
\def\@arrayskip{\ifold\baselineskip\z@\lineskip\z@
  \else
  \baselineskip\minarrayskip\lineskip1\baselineskip\fi}
\def\@arrayclassz{\ifcase \@lastchclass \@acolampacol \or
\@ampacol \or \or \or \@addamp \or
\@acolampacol \or \@firstampfalse \@acol \fi
\edef\@preamble{\@preamble
\ifcase \@chnum
  \hfil$\relax\arraymode\@sharp$\hfil
  \or $\relax\arraymode\@sharp$\hfil
  \or \hfil$\relax\arraymode\@sharp$\fi}}
\def\@array[#1]#2{\setbox\@arstrutbox=\hbox{\vrule
  height\arraystretch \ht\strutbox
  depth\arraystretch \dp\strutbox
width\z@}\@mkpream{#2}\edef\@preamble{\halign \noexpand\@halignto
\bgroup \tabskip\z@ \@arstrut \@preamble \tabskip\z@ \cr}%
\let\@startpbox\@@startpbox \let\@endpbox\@@endpbox
\if #1t\vtop \else \if#1b\vbox \else \vcenter \fi\fi
\bgroup \let\par\relax
\let\@sharp##\let\protect\relax
\@arrayskip\@preamble}
\def\eqnarray{\stepcounter{equation}%
           \let\@currentlabel=\theequation
           \global\@eqnswtrue
           \global\@eqcnt\z@
           \tabskip\@centering              
           \let\\=\@eqncr
           $$%
         \halign to \displaywidth  \bgroup
          \eqnumphantom \@eqnsel
   \hskip\@centering                               
 $\displaystyle  \tabskip\z@ {##}$%
 &\global\@eqcnt\@ne \hskip 2\arraycolsep
      $ \displaystyle  \arraymode{##}$\hfil
 &\global\@eqcnt\tw@ \hskip 2\arraycolsep
      $\displaystyle\tabskip\z@{##}$\hfil
      \tabskip\@centering
 &{##}\tabskip\z@\cr}
\makeatother

\newcounter{mo}

\newcounter{bk}

\newcommand{\Si}{\Sigma}
\newcommand{\tr}{{\rm tr}}

\newcommand{\Ad}{{\rm Ad}}
\newcommand{\ti}[1]{\tilde{#1}}
\newcommand{\om}{\omega}
\newcommand{\Om}{\Omega}
\newcommand{\de}{\delta}
\newcommand{\al}{\alpha}
\newcommand{\te}{\theta}
\newcommand{\vth}{\vartheta}
\newcommand{\be}{\beta}
\newcommand{\la}{\lambda}

\newcommand{\ep}{\epsilon}

\newcommand{\G}{\Gamma}
\newcommand{\ka}{\kappa}

\newcommand{\ga}{\gamma}

\newcommand{\si}{\sigma}

\newcommand{\mat}[4]{\left(\begin{array}{cc}{#1}&{#2}\\{#3}&{#4}
\end{array}\right)}

\def\bea{\begin{eqnarray}\new\begin{array}{cc}}
\def\ee{\end{array}\end{eqnarray}}

\newcommand{\beq}[1]{\begin{equation}\label{#1}}
\newcommand{\eq}{\end{equation}}
\newcommand{\beqn}[1]{\begin{small} \begin{eqnarray}\label{#1}}
\newcommand{\eqn}{\end{eqnarray} \end{small}}
\newcommand{\p}{\partial}
\def\sq2{\sqrt{2}}
\newcommand{\di}{{\rm diag}}
\newcommand{\oh}{\frac{1}{2}}

\def\sl2{{\rm sl}(2, {\mathbb C})}
\def\SLN{{\rm SL}(N, {\mathbb C})}

\def\SLT{{\rm SL}(2, {\mathbb C})}
\def\GLT{{\rm GL}(2, {\mathbb C})}

\def\f1#1{\frac{1}{#1}}

\newcommand{\bp}{\bar{\partial}}
\newcommand{\bz}{\bar{z}}

\newcommand{\bA}{\bar{A}}
\newcommand{\bG}{\bar{G}}

\def\mC{{\mathbb C}}
\def\mZ{{\mathbb Z}}
\def\mR{{\mathbb R}}

\def\frak{\mathfrak}
\def\gb{{\frak b}}
\def\gc{{\frak c}}
\def\gg{{\frak g}}
\def\gp{{\frak p}}

\def\gn{{\frak n}}

\def\gM{{\frak M}}

\def\gk{{\frak k}}

\def\gh{{\frak h}}

\def\gu{{\frak u}}
\def\gx{{\frak x}}

\def\bfe{{\bf e}}

\def\bfS{{\bf S}}

\def\bfX{{\bf X}}

\def\bfT{{\bf T}}

\def\clB{\mathcal{B}}
\def\clC{\mathcal{C}}
\def\clD{\mathcal{D}}

\def\clG{\mathcal{G}}

\def\clO{\mathcal{O}}
\def\clH{\mathcal{H}}
\def\clK{\mathcal{K}}

\def\clM{\mathcal{M}}
\def\clN{\mathcal{N}}
\def\clP{\mathcal{P}}
\def\clQ{\mathcal{Q}}

\def\clX{\mathcal{X}}

\def\clZ{\mathcal{Z}}

\def\bag2{{\bf g_2}}
\def\bas8{{\bf so(8)}}

\def\sr2{\sqrt{2}}

\def\f1#1{\frac{1}{#1}}

\newtheorem{predl}{Proposition}[section]

\newtheorem{rem}{Remark}[section]

\def\Ad{{\rm Ad}}
\def\Bun{{\rm Bun}}
\def\End{{\rm End}}
\def\rank{{\rm rank}}

\def\Res{{\rm Res}}

\def\Fl{{\rm Fl}}

\begin{document}
 \begin{flushright}
 ITEP-TH-14/20\\
 IITP-TH-10/20
 \end{flushright}
\vspace{3mm}

 \begin{center}
{\LARGE Generalizations of parabolic Higgs  bundles, real structures}
\\ \vspace{4mm}
{\LARGE and integrability}\\
 \vspace{10mm}
  {{\large A. Levin}$^{\,\natural\,\,\flat}$\ \ \ \
  {\large M. Olshanetsky}$^{\,\flat\,\S}$\ \ \ \ {\large A. Zotov}$^{\,\diamondsuit\,\flat\, \natural}$}\\
   \vspace{7mm}
  \vspace{2mm}$^\flat$ -
  {\sf ITEP of NRC ''Kurchatov Institute'',\\ B. Cheremushkinskaya, 25, Moscow, 117259, Russia}\\
   \vspace{2mm}$^\S$ - {\sf Institute for Information Transmission Problems RAS (Kharkevich Institute),
 \\  Bolshoy Karetny per. 19, Moscow, 127994,  Russia}\\
 \vspace{2mm}$^\natural$ - {\sf National Research University Higher School of Economics, Russian Federation,\\
 NRU HSE,
 Usacheva str. 6,  Moscow, 119048, Russia}\\
  \vspace{2mm} $^\diamondsuit$ -
 {\sf Steklov Mathematical Institute of Russian
Academy of Sciences,\\ Gubkina str. 8, Moscow, 119991,  Russia
 }\\

 \vspace{4mm}
 {\footnotesize Emails: alevin2@hse.ru, olshanet@itep.ru,
 zotov@mi-ras.ru}\\
 \end{center}

 \begin{abstract}
We introduce a notion of quasi-antisymmetric Higgs $G$-bundles over curves with marked points.
They are endowed with additional structures,
 which replace the parabolic structures at marked points in the parabolic Higgs bundles. The latter
means that the coadjoint orbits are attached to the marked points.
 The moduli spaces of parabolic Higgs bundles are the phase spaces of complex completely integrable systems.
 In our case the coadjoint orbits are replaced by the cotangent bundles over some special symmetric
 spaces in such a way that the moduli space of the modified Higgs bundles are still phase
 spaces of complex completely integrable systems.
 We show that the moduli space of the parabolic Higgs bundles is the
  symplectic quotient of the moduli space of the  quasi-antisymmetric Higgs bundle with respect to the action of product of Cartan subgroups.
  Also, by changing the symmetric spaces we introduce
 quasi-compact and quasi-normal Higgs bundles. Then the fixed point sets of real involutions
  acting on their  moduli spaces are the phase spaces of real completely
  integrable systems.
  Several examples are given including integrable extensions of the ${\rm SL}(2)$ Euler-Arnold top, two-body
  elliptic Calogero-Moser system and the rational  ${\rm SL}(2)$  Gaudin system together with its real reductions.
 \end{abstract}


\newpage

\footnotesize \tableofcontents \normalsize

\section{Introduction and summary}

\setcounter{equation}{0}

Let $G^\mC$ be a simple complex Lie group, $\Si_g$ is a Riemann curve
of genus $g$ with a canonical class $\ka$ and $E$ is a vector $G^\mC$-bundle over $\Si_g$.
The Higgs bundle is pair $(d_{\bA},\Phi)$, where $d_{\bA}$ is a holomorphic structure on
the bundle $E$, and $\Phi$ is the Higgs field -- a section of the bundle $End(E)\otimes\ka$.
The Higgs bundles are a symplectic spaces. The symplectic quotient with respect to the action of
gauge group $\clG(\Si_g,G^\mC)=\{C^\infty(\Si_g)\to G^\mC\}$ is the moduli space of Higgs bundles
 $\clM(\Si_g,G^\mC)$.

If the curve is smooth then the moduli space of the Higgs bundles are the phase spaces
of classical complex integrable systems called the Hitchin systems \cite{Hi}. If at some points
 $(x_1,\ldots,x_n)$ of $\Si_{g,n}$
 the gauge group is reduced to the Borel subgroup, or more generally
to  parabolic subgroups \cite{Ko,Si} then the corresponding  Higgs bundle is called
parabolic. This construction is equivalent to attachment to the marked points the coadjoint
orbits
\footnote{In Appendix we prove this equivalence in general setting.
In this paper we use both constructions.}.
The moduli space $\clM_{par}(\Si_{g,n},G^\mC)$ of parabolic Higgs bundles is also
a phase spaces of an integrable systems \cite{BKV,Ne}.  This approach to integrable systems
allows one to construct the Lax operators,
integrals of motion, action-angle variables, classical r-matrices and so on. Many integrable systems including
the Calogero-Moser systems, the Euler-Arnold elliptic tops and the Gaudin type systems can be described in this way.

The moduli space of the Higgs bundles over smooth curves is a symplectic quotient of the moduli
space of parabolic bundles with respect to the action of some finite-dimensional group $\ti G_0$
\beq{fol}
\clM(\Si_g,G^\mC)=\ti G_0\setminus\setminus\clM_{par}(\Si_{g,n},G^\mC)\,,\qquad \ti G_0=\prod_aG^\mC\,,
\eq
see (\ref{fga}).


In this paper we modify the parabolic Higgs bundles in such a way that
their moduli spaces  also become the phase spaces of integrable systems (real or complex). Namely,
 we replace the Borel subgroups $B^\mC$  with some special reductive subgroups. If these subgroups
are some special complex group $U^\mC\subset G^\mC$ (for example, for $G^\mC=\SLN$,  $U^\mC=$SO$(N,\mC)$) then we come
 to the complex integrable systems as before.

Another possibility is to choose this subgroup to be the maximal compact subgroup or the normal subgroup of
$G^\mC$.
 The resulting  moduli spaces are real varieties. To come to integrable system we
use  involutions defined on the moduli space of Higgs bundles proposed in \cite{BS}. The
 fixed point set of the involutions are the phase spaces of classical real completely integrable
 systems.

First examples of such system were proposed in \cite{FP}.   It is the Calogero-Sutherland system with the two types of spin variables.
In our previous work \cite{KLOZ} we defined these systems using this new type of the Higgs bundles.
In this paper we develop a general approach to this Higgs bundles and consider new
examples.

\noindent {\bf  Outline of the paper}

Our goal is to investigate the moduli spaces of these modifications of the parabolic Higgs bundles
and to construct completely integrable systems. For this purpose
 we replace the co-adjoint orbits attached to the marked points in the parabolic Higgs
bundles by symplectic manifolds in such a way that the corresponding moduli spaces are
still the phase spaces of integrable systems.
These symplectic spaces  are cotangent bundles to some special symmetric spaces defined as
follows.

Consider a maximal compact subgroup  $C$ of $G^\mC$ (and its normal form  $G^\mR$
\cite{He}).
For example, for $G^\mC=\SLN$ the maximal compact subgroup is SU$(N)$ and $G^\mR$ is
SL$(N,\mR)$. The subgroups $C$ and $G^\mR$ are fixed point sets of the
commuting anti-holomorphic involutive automorphisms
$\rho$ and $\si$ of $G^\mC$
$$
\rho(C)=C\,,~~\si(G^\mR)=G^\mR\,.
$$
 Let $U$ be a maximal compact subgroup of $G^\mR$:
 $$
 U=\{g\in G^\mR\,|\,\rho(g)=g\}\,.
 $$
 Or, equivalently,
 $$
 U=\{g\in C\,|\,\si(g)=g\}\,.
 $$
  For the group
SL$(N,\mC)$ the subgroup $U$ is SO$(N,\mR)$.
Let $U^\mC$ be the complexification of the real group $U$. It is the fixed
point set in $G^\mC$ of the involutive automorphism $\te=\si\circ\rho$
$$
U^\mC=\{g\in G^\mC\,|\,\te(g)=g\}\,.
 $$
For $G^\mC=\SLN$ the corresponding subgroup is $U^\mC=$SO$(N,\mC)$.

Define five types of the coset spaces:
$$
\begin{array}{lll}
  {\bf I.} &  ~~\clX_I=C\backslash G^\mC&  ~~\rho\,(C)=C \\
  {\bf II.}& ~~\clX_{II}=G^\mR\backslash G^\mC& ~~ \si\,(G^\mR)=G^\mR \\
  {\bf III.}& \clX_{III}=U\backslash C&  ~~\si\,(U)=U \\
  {\bf IV.}&  ~~\clX_{IV}=U \backslash  G^\mR&~~ \rho\,(U)=U \\
  {\bf V.}& ~~\clX_{V}=U^\mC\backslash G^\mC& ~~ \te\,(U^\mC)=U^\mC
\end{array}
$$
\begin{center}
\texttt{Table 1}
\end{center}
They are symmetric spaces \cite{Be,He}. It means that their stationary subgroups are  the fixed point sets of the corresponding involutive automorphisms.
 The symmetric spaces of types I, III and IV  are Riemannian manifolds, while the spaces of type II and V are pseudo-riemannian manifolds. The action of involutions leads to
 the following interrelations between symmetric spaces:
 \begin{subequations}\label{001}
  \begin{align}
& \clX_{IV}=U \backslash G^\mR\stackrel{\si}{\hookrightarrow}\clX_I=C \backslash G^\mC\,,\\
 &\clX_{IV}=U \backslash G^\mR\stackrel{\si}{\hookrightarrow}\clX_V=U^C \backslash G^\mC\,,
\end{align}
\end{subequations}
where the arrows mean the embedding as the fixed  points set.
Similarly,
 \begin{subequations}\label{002}
  \begin{align}
&\clX_{III}=U\backslash C\stackrel{\rho}{\hookrightarrow}\clX_{II}=G^{\mR}\backslash G^\mC\,,\\
 & \clX_{III}=U\backslash C\stackrel{\rho}{\hookrightarrow}\clX_{V}=U^C \backslash G^\mC\,.
\end{align}
\end{subequations}

Let $\clD=(x_1,\ldots,x_n)$ be the set of the marked points on the curve $\Si_{g,n}$.
Recall that \emph{the parabolic structure} of the Higgs
bundles over $\Si_{g,n}=\Si_g\setminus\clD$ means that we attach to the marked points coadjoint
 $G^\mC$-orbits $\clO_a$ $(a=1,\ldots,n)$.
As we mentioned above the moduli space of the parabolic Higgs bundles $\clM_{par}(\Si_{g,n},G^\mC)$
are  phase spaces of completely integrable system. The Poisson commuting
independent integrals of motion are constructed by means of the generalized Beltrami
differentials on $\Si_{g,n}$. The number of integrals is independent of structures
attached to the marked points.
It turns out that it is
equal to $\oh\dim\,(\clM_{par}(\Si_{g,n},G^\mC)$. This provides the Liouville integrability
of the systems related to the parabolic Higgs bundles.

 There is another way to define  the parabolic Higgs bundles.
 Instead of attaching the coadjoint orbits to the marked points
 (the symplectic charges) one can reduce the gauge group (in an appropriate way).
 For the parabolic bundles the gauge group takes values in Borel subgroups
 at the marked points. We call
 this construction the {\bf $\be$-model}, while the first construction is called the {\bf $\al$-model}
 \footnote{
 These two descriptions of charges are known in gauge theories. It is "electric", or
  "magnetic" points of view. Unfortunately, we did not find the proof of their equivalence in the literature. For this reason we  give it in Appendix A.}.

We define the following modifications of the parabolic Higgs bundles that eventually leads to the
integrability.

{\bf The quasi-antisymmetric structure} of the Higgs
bundles means that we attach to the marked points the cotangent bundles
$T^*(\clX_{V,a})=T^*(U^{\mC}_a\backslash G^\mC)$. We prove that the moduli spaces
 $\clM_V(\Si_{g,n},G^\mC)$ of the quasi-antisymmetric Higgs bundles
 are the phase spaces of complex completely integrable systems.
The moduli space of the parabolic Higgs bundles $\clM_{par}(\Si_{g,n},G^\mC)$
is a symplectic quotient of the moduli space the quasi-antisymmetric Higgs bundles $\clM_{V}(\Si_{g,n},G^\mC)$ with respect to the action of a finite-dimensional
group (see $\clH$ in (\ref{clh})):
\beq{asp}
\clM_{par}(\Si_{g,n},G^\mC)=\clH\setminus\!\setminus\clM_{V}(\Si_{g,n},G^\mC)\,.
\eq
 The latter means that the systems of type V are integrable extensions of the parabolic Hitchin systems.
 The coordinates on $\clM_{par}(\Si_{g,n},G^\mC)$ play the role of collective
 coordinates on the phase space $\clM_{V}(\Si_{g,n},G^\mC)$.

The real integrable systems arise in the following way.
 Assume that $\Si_{g,n}$ admits an anti-holomorphic involution
$\jmath\,:\,\Si_{g,n}\to\Si_{g,n}$.
In a neighborhood of  fixed points of this action
one can find a local
coordinate $z$ such that the involution $\jmath$ is given by the complex conjugation
$\jmath(z)=\bz$ \cite{GH}. The fixed point set can be seen as a union
$\Si^0=\cup S^1_a$ of copies of the unit circle $S^1$ embedded in $\Si_{g,n}$.
Assume that the set $\clD$ of the marked points is invariant with respect to the involution
$\jmath(\clD)=\clD$.
 Following \cite{BS} define two involutions on the moduli space of Higgs
bundles $\clM(\Si_g,G^\mC)$
\footnote{In \cite{BS} they are  denoted as $\iota_2$ and $\iota_3$.}
\beq{io}
\iota^\rho( d_{\bA}),\Phi)=(\jmath^*\rho (d_{\bA}),-\jmath^*\rho(\Phi))\,,~~
\,\,
   \iota^\si( d_{\bA}),\Phi)=(\jmath^*\si (d_{\bA}),\jmath^*\si(\Phi))\,.
\eq
The fixed point set of $\iota^\si$ is the Higgs $G^\mR$-bundle over $\Si^0$ and the
fixed point set of $\iota^\rho$ is the Higgs $C$-bundle over $\Si^0$.
As an intermediate step
we construct  two types of Higgs bundles using the cotangent bundles
 to symmetric spaces of type I or type II.

\paragraph{Type I.} Attach  to the marked points the cotangent bundles
$T^*(\clX_{I,a})=T^*(C_a \setminus G^\mC)$. We call these Higgs bundles {\bf quasi-compact}, because
at the marked points
the gauge group of the bundle is reduced to the maximal compact subgroups $C_a$.
The moduli space of the quasi-compact Higgs bundles $\clM_I(\Si_{g,n},G^\mC)$ is a real space.
Its dimension  exceeds the number of integrals of motion. To come
to integrable systems we use the involution $\iota^\si$ (\ref{io}) acting on $\clM_I(\Si_{g,n},G^\mC)$. It follows from (\ref{001}a) that the fixed set of involution
$\iota^\si$  is the moduli space $\clM_{IV}(\Si^0_n,G^\mR)$
of the $G^\mR$ Higgs bundles over $\Si^0$,
 where the cotangent bundles attached to the marked points
 are $T^*(\clX_{IV,a})=T^*(U_a\setminus G^\mR)$.
 Acting by the same involutions on the moduli space of the antisymmetric Higgs bundles
 $\clM_V(\Si_{g,n},G^\mC)$ and using (\ref{001}b) we come to the same fixed point set  $\clM_{IV}(\Si_{g,n},G^\mC)$ in $\clM_V(\Si_{g,n},G^\mC)$. Thus we have the embeddings
 \beq{io1}
 \clM_{IV}(\Si^0_n,G^\mR)\stackrel{\iota^\si}{\hookrightarrow} \clM_I(\Si_{g,n},G^\mC)\,,~~
 \clM_{IV}(\Si^0_n,G^\mR)\stackrel{\iota^\si}{\hookrightarrow} \clM_V(\Si_{g,n},G^\mC)\,.
 \eq
We prove that moduli space  $\clM_{IV}$ is a phase space of a real completely integrable system.

\paragraph{ Type II.} Replace in the above construction
the cotangent bundles $T^*(\clX_{I,a})$
  with the cotangent bundles $T^*(\clX_{II,a})=T^*(G^\mR_a\setminus G^\mC)$.
 We call these Higgs bundles \emph{quasi-normal} since $G^\mR_a$  is the normal
 subgroup of $G^\mC$.
The moduli space of the quasi-normal Higgs bundles $\clM_{II}(\Si_{g,n},G^\mC)$ is too large
to be a phase space of completely integrable system. The integrable systems
 arise on the fixed point set of the involution $\iota^\rho$ (\ref{io})
 acting on  $\clM_{II}(\Si_{g,n},G^\mC)$ and on $\clM_{V}(\Si_{g,n},G^\mC)$. Using  (\ref{002}a) and (\ref{002}b) we come
 to the  moduli space $\clM_{III}(\Si^0_n,C)$ of the Higgs $C$-bundles over $\Si^0$ embedded in
 the moduli spaces $\clM_{II}$ and $\clM_{V}$
 \beq{io2}
 \clM_{III}(\Si^0_n,C)\stackrel{\iota^\rho}{\hookrightarrow}\clM_{II}(\Si_{g,n},G^\mC)\,,~~
 \clM_{III}(\Si^0_n,C)\stackrel{\iota^\rho}{\hookrightarrow}\clM_{V}(\Si_{g,n},G^\mC) \,.
 \eq
$\clM_{III}$  is also a phase space of a real completely integrable system.

In the absence of the marked points  both involutions and their fixed point sets
as the phase space of real integrable systems were considered in \cite{BS,BS1}

Schematically, we have the  following interrelations between  the moduli spaces:
 \beq{dia}
\xymatrix{
  \clM_{II}  &
   & \clM_V\ar[d]_{\clH\setminus\!\setminus}   & & \clM_I          \\
 & \ar[ur]^{\iota^\rho}\clM_{III}\ar[ul]^{\iota^\rho} &\clM_{par} & \ar[ur]^{\iota^\si}    \clM_{IV}      \ar[ul]^{\iota^\si}
  }
 \eq
 Here the passage from $\clM_V$ to $\clM_{par}$ means the symplectic reduction
 (\ref{asp}), while the other arrows denote the embeddings as the fixed point sets.
While the moduli spaces $\clM_I$ and $\clM_{II}$ play a supporting role, the remaining
moduli spaces are the phase spaces of integrable systems.

There is a universal generalization of the parabolic Higgs bundles, which we call {\bf type 0
Higgs bundles}.
For these bundles we attach to the marked points the
cotangent bundles $T^*G^\mC$. The moduli space $\clM_{0}(\Si_{g,n},G^\mC)$ of the type 0 Higgs bundles
allows one to define the moduli spaces of the types I,II and V Higgs bundles by
symplectic reductions with respect to action of the groups $G_I$,  $\,G_{II}$ and $G_V$ (\ref{fgg}):
$$
\xymatrix{
  &\ar[dl]_{G_{II}\setminus\!\setminus} \clM_0 \ar[d]_{G_V\setminus\!\setminus} \ar[dr]^{G_I\setminus\!\setminus} &\\
  \clM_{II}&\clM_V&\clM_I
        }
$$
Similarly to the parabolic bundles the Higgs bundles with  the cotangent bundles
attached to the marked points after symplectic reductions become the Higgs bundles over
the smooth curves (see (\ref{fol}))
\beq{ext}
\clM_J(\Si_{g,n},G^\mC)\stackrel{\ti G_0 \setminus\!\setminus} {\longrightarrow}\clM(\Si_g,G^\mC)\,,~~(J=0,I,II,V)\,,
\eq
where $\ti G_0$ is from (\ref{fol}).
It is possible to attach to the marked points the cotangent bundles to the Euclidean symmetric spaces
 of zero curvature. The corresponding systems are analogues of rational integrable systems, while
the systems related to  $\clM_{III}$ and $\clM_{IV}$ are analogues of the hyperbolic and
trigonometric systems. We don't consider the rational case here.

\bigskip
\noindent
\paragraph{Organization of the paper.}

In Section 2 we construct the Higgs bundles over curves with marked points of types $J=0,I,II,V$
based on the $\al$ and $\be$ models.
 For the bundles of type V we construct the integrals of motion and prove the
  complete integrability of the systems.

In Section 3  we consider the real involutions for the systems of type V.
The fixed point sets of the involutions are the real integrable systems of types III and IV.
We consider the mixing of these systems with the systems of the parabolic type singularities.


In Section 4 we describe the Higgs bundles using the double coset construction of the moduli
space of holomorphic bundles. This construction allows one to define topologically non-trivial
Higgs bundles described by their characteristic classes.
The construction of the parabolic bundles is related to the affine flag
varieties \cite{PS}. For the other types of the Higgs bundles the similar role is played by the  affine symmetric spaces.

In Section 5 we consider three examples
of quasi-antisymmetric and real Higgs bundles. First we consider a system on $\Si_{0,n}$
$\,(\mC P^1$ with $n$ marked points). It is an extension
of the rational Gaudin system corresponding to the symplectic reduction (\ref{asp}). A special configuration of the marked points
has two involutions corresponding to the real integrable systems of type III and IV.
 Next example is the system of type V on  $\Si_{1,1}$  (an elliptic curve with one
marked point) and group $G^\mC=$SL$(2,\mC)$.
The trivial bundles lead to the generalization of the two-body
elliptic  Calogero-Moser system. For the quasi-antisymmetric Higgs bundles the coupling constant
 acquires internal degree of freedom with explicit hamiltonian dynamics.
For non-trivial bundles  the moduli space of the quasi-antisymmetric Higgs
 bundles $\clM_V$ is isomorphic to
the cotangent bundle $T^*($SL$(2,\mC)/\mC^*)$, ($\dim\,(\clM_V)=4)$.
We construct the Lax operator and two commuting integrals of motion and establish the connection
of this model to the $\SLT$ Euler-Arnold top coming from the diagram (\ref{asp}).

In the concluding Section we discuss interpretation of the presented results in terms of the
twisted $N=4$, $d=4$ SUSY Yang-Mills theory \cite{KW,GW,GW2}. In this context the quasi-antisymmetric Higgs bundles are related to the surface operators that apparently break  $N=4$ supersymmetry to $N=2$.


In the Appendix we prove the equivalence of $\al$ and $\be$ models descriptions of the Higgs bundles over
multipoint curves.
Then we provide the necessary information regarding symmetric spaces and flag varieties.
Finally, we describe the symplectic structures on the coadjoint orbits and on the
cotangent bundles to symmetric spaces by the symplectic reduction from the cotangent bundle $T^*G^\mC$
and define the Darboux coordinates on cotangent bundles to symmetric spaces.

\noindent \paragraph{Acknowledgements}
The work of M. Olshanetsky was funded by the Russian Science Foundation (Grant No. 16-12-10344).
 The work of A. Levin was
partially supported partially supported by Laboratory of Mirror Symmetry NRU HSE, RF Government grant, ag. 14.641.31.0001.
The research of A. Zotov was supported
by the HSE University Basic Research Program and by the Young Russian Mathematics award.

\section{General construction}

\setcounter{equation}{0}

\subsection{Vector bundles}
Let $G^\mC$ be a complex simple Lie group, $\Si_g$ is a Riemann curve of genus $g$ and $\clP(G^\mC)$ is the principle $G^\mC$ - bundle over $\Si_g$. Define associated vector bundle $E(G^\mC)=\clP\times_{G^\mC}V$, where $V$
is a $G^\mC$ module.
Let $d_{\bA}$ be a holomorphic structure on the bundle $E(G^\mC)$. The
 moduli space of the holomorphic bundles $Bun(\Si_g,G^\mC)$  is the quotient
\beq{bud}
Bun(\Si_g,G^\mC)=\{d_{\bA}\}/\clG(\Si_g,G^\mC)\,,
\eq
where
\beq{sgg}
\clG(\Si_g,G^\mC)=\{C^\infty(\Si_g)\to G^\mC\}
\eq
 is the gauge group.
It acts on $d_{\bA}$ by the affine transformations
\beq{fgt}
d_{\bA}\to f d_{\bA}f^{-1}\,.
\eq
Here we consider the topologically trivial bundles.
 The non-trivial bundles will be introduced in Section 5.

 The dimension of $Bun(\Si_g,G^\mC)$  can be extracted from the Riemann-Roch theorem.
 For stable bundles we have
 \beq{fd}
\dim\,Bun(\Si_g,G^\mC)=(g-1)\dim\,G^\mC\,.
\eq

Consider a multipoint curve  $\Si_{g,n}$ with $n$ marked points $\clD=(x_1,\ldots,x_n)$.
In the $\al$-model we attach to them different data, which are not necessarily holomorphic. The known example is
the holomorphic bundle with the\emph{ parabolic structure}. In this case we
attach the flag varieties $Fl^\mC_a=B_a^\mC \setminus G^\mC$, where $B^\mC_a$ are Borel subgroups
of $G^\mC$. The moduli space of holomorphic bundles with parabolic structure is defined as
the quotient space
$$
Bun_{par}(\Si_{g,n},G^\mC)=\{(d_{\bA}\}\,,\,\prod Fl^\mC_a)\}/\clG(\Si_{g},G^\mC)\,.
$$
 From (\ref{fd}) and (\ref{df}) we have
\beq{fd1}
\dim\,Bun_{par}(\Si_{g,n},G^\mC)=(g-1)\dim\,G^\mC+n\dim\,Fl^\mC=(g-1)\dim\,G^\mC+n\sum_{j=1}^l(d_j-1)\,,
\eq
where $d_j$ are the orders of the invariants of the algebra $\gg^\mC$ (\ref{dj}) and
$l=\rank(\gg^\mC)$.

It follows from Proposition A.1 that $Bun_{par}$ can be equivalently defined in the $\be$-model by
 reducing the gauge group to the subgroup
\beq{cpar}
\clG_{par}(\Si_{g,n},G^\mC):=(\clG(\Si_g,G^\mC)|_{z=x_a}=B^\mC_a)\,.
\eq
Then
\beq{mpar}
Bun_{par}(\Si_{g,n},G^\mC)=\{d_{\bA}\}/\clG_{par}(\Si_{g,n},G^\mC)\,,
\eq
where (in the notations of the Proposition A.1) we identify $X=\{d_{\bA}\}$, $G=\clG(\Si_g,G^\mC)$ and $H=\clG_{par}(\Si_{g,n},G^\mC)$
so that $H\setminus G=\prod_a\Fl_a^\mC$.
Let
\beq{g0}
G_0=\prod_a G^\mC\,.
\eq
Since the group $G^\mC$ acts transitively on the flag variety we find that
\beq{ps}
Bun(\Si_g,G^\mC)=G_{0}\setminus Bun_{par}(\Si_{g,n},G^\mC)\,,
\eq
where the leave of the of the projection $\pi\,:\,Bun_{par}(\Si_{g,n},G^\mC)\to Bun(\Si_g,G^\mC)$  is $\prod_a Fl_a$.

We are going to modify this construction in the following ways. Replace the flag varieties with
 four different types of quotient spaces.

\paragraph{Type 0.}
Consider two isomorphic representations $V$ and $V_a$ of $G^\mC$, where $V_a$ is the fiber over
the marked point $x_a$ and $V$ is fixed. Denote by
$Isom(V_a, V)$ the space of $G^\mC$-isomorphisms between $V_a$ and  $V$. This space
is a principal homogeneous space over $G^\mC$.

For  curves  with  marked points  we replace the moduli
space $Bun(\Si_g,G^\mC)$ by the moduli space
 of holomorphic bundles with the trivializations $g_a: V_{a}\to V$
of fibers at the marked points
\beq{t0}
Bun_0(\Si_{g,n},G^\mC)=\Big\{d_{\bA}\,,\,\prod_{a=1}^nIsom(V_a,V)\Big\}/\clG(\Si_g,G^\mC)\,.
\eq
 Then
 \beq{d0}
 \dim\,Bun_0(\Si_{g,n},G^\mC)=(g-1+n)\dim\,G^\mC\,.
\eq
In the $\be$-model the gauge group is
\beq{G0}
\clG_0(\Si_g,G^\mC)=\{f\in \clG(\Si_g,G^\mC)\,|\,f(x_a)=Id\,,~(a=1,\ldots,n)\}\,,
\eq
and $Bun_0(\Si_{g,n},G^\mC)$ is defined as
\beq{t00}
Bun_0(\Si_{g,n},G^\mC)=\{d_{\bA}\}/\clG_0(\Si_{g,n},G^\mC)\,.
\eq
 As above we have the projection
 $\pi\,:\,Bun_0(\Si_{g,n},G^\mC)\to Bun(\Si_g,G^\mC)$ and
\beq{fb}
Bun(\Si_g,G^\mC)=G_0\setminus Bun_0(\Si_{g,n},G^\mC) \,,~~G_0=\prod_aG^\mC\,.
\eq

\paragraph{Type I.}
Let $C$ be the maximal compact subgroup of $G^\mC$, defined by the involutive automorphism $\rho$
(\ref{rho}).
In the $\al$-model we attach to the marked points the symmetric spaces
 $\clX_{I,a}=C_a\setminus G^\mC$. By definition, the moduli space of bundles with \emph{ quasi-compact structure}
  is the quotient space
 \beq{bI}
Bun_I(\Si_{g,n},G^\mC)=\{(d_{\bA}\}\,,\,\prod \clX_{I,a})\}/\clG(\Si_{g},G^\mC)\,.
\eq
Equivalently, in the $\be$-model
$$
Bun_I(\Si_{g,n},G^\mC)=\{(d_{\bA}\}/\clG_I(\Si_{g,n},G^\mC)\,,
$$
where
\beq{g1}
\clG_I(\Si_{g,n},G^\mC)=\{f\in\clG(\Si_{g},G^\mC)\,|\,f(x_a)\in C_a)\,,~ (a=1,\ldots,n)\}\,.
\eq
Note that the holomorphic structure is broken at the marked points, and $Bun_I(\Si_{g,n},G^\mC)$
is a real space.
From (\ref{dgx}) we find
\beq{dI}
\dim_\mR\,Bun_I(\Si_{g,n},G^\mC)=(g-1)\dim_\mR\,G^\mC+n\dim_\mR\,X_I=
(g-1)\dim_\mR\,G^\mC+n \sum_{j=1}^l( 2d_j-1)\,.
\eq
Since $\clX_I=C\setminus G^\mC$ and the left action commutes with the right action of
$\clG_I(\Si_{g,n},G^\mC)$ we have
\beq{fb1}
Bun_I(\Si_{g,n},G^\mC)=G_I\setminus Bun_0(\Si_{g,n},G^\mC)\,,~~G_I=\prod_a C_a\,.
\eq
The group $G^\mC$ acts transitively on the symmetric space $\clX_I$. Thereby, similarly to (\ref{fb})
\beq{fb10}
Bun(\Si_g,G^\mC)=G_0\setminus Bun_I(\Si_{g,n},G^\mC)\,,
\eq
where $G_0$ is (\ref{g0}).

\paragraph{Type II.}
 Let $G^\mR$ be the normal subgroup of $G^\mC$ defined by the
involutive automorphism $\si$ (\ref{si}).
 Attach to the marked  points the symmetric spaces
$\clX_{II,a}=G^\mR_a \setminus G^\mC$.
The moduli space of bundles with \emph{the quasi-normal structure} is the quotient space
 \beq{bII}
Bun_{II}(\Si_{g,n},G^\mC)=\{(d_{\bA}\}\,,\,\prod \clX_{IIa})\}/\clG(\Si_{g},G^\mC)\,,
\eq
or equivalently,
$$
 Bun_{II}(\Si_{g,n},G^\mC)=\{(d_{\bA}\}/\clG_{II}(\Si_{g,n},G^\mC)\}\,,
$$
where
\beq{g2}
\clG_{II}(\Si_{g,n},G^\mC)=\{f\in\clG(\Si_{g},G^\mC)\,|\,f(x_a)\in G^\mR_a\}\,.
\eq
Again, $Bun_{II}(\Si_{g,n},G^\mC)$ is a real space and its dimension coincides with $\dim\,Bun_I$:
\beq{dII}
\dim_\mR\,Bun_{II}(\Si_{g,n},G^\mC)=(g-1)\dim_\mR\,G^\mC+n\dim_\mR\,X_{II}=
\dim_\mR\,Bun_{I}(\Si_{g,n},G^\mC)\,,
\eq
(see (\ref{dgr})).
In addition, as in (\ref{fb1}) we have
\beq{fb2}
Bun_{II}(\Si_{g,n}G^\mC)=G_{II}\setminus Bun_{0}(\Si_{g,n},G^\mC)\,,~~G_{II}=\prod_a G^\mR_a\,,
\eq
and
\beq{fb20}
Bun(\Si_g,G^\mC)=G_0\setminus Bun_{II}(\Si_{g,n},G^\mC) \,.
\eq

\paragraph{Type V.}
 Let $U^\mC$ be the complex subgroup of $G^\mC$ defined by the
involutive automorphism $\te=\si\rho$.
Attach to the marked  points the symmetric spaces
$\clX_{V,a}=U^\mC_a \setminus G^\mC$.
The moduli space of bundles with \emph{quasi-antisymmetric structure} is the quotient space
$$
Bun_{V}(\Si_{g,n},G^\mC)=\{(d_{\bA}\,,\,\prod \clX_{Va})\}/\clG(\Si_{g},G^\mC)\,,
$$
or equivalently,
$$
 Bun_{V}(\Si_{g,n},G^\mC)=\{d_{\bA}\}/\clG_{V}(\Si_{g,n},G^\mC)\,,
$$
where
\beq{g5}
\clG_{V}(\Si_{g,n},G^\mC)=\{f\in\clG(\Si_{g},G^\mC)\,|\,f(x_a)\in U^\mC_a\}\,.
\eq
The space $ Bun_{V}(\Si_{g,n},G^\mC)$ is complex. Its dimension is equal to
\beq{bV}
\dim_\mC\,Bun_{V}(\Si_{g,n},G^\mC)=(g-1)\dim_\mC\,G^\mC+n\dim_\mC\,\clX_{V}=
\eq
$$
=(g-1)\dim_\mC\,G^\mC+n \sum_{j=1}^ld_j\,,
$$
(see (\ref{d5})).
Finally,
\beq{fb5}
Bun_V(\Si_{g,n},G^\mC)= G_V \setminus Bun_0(\Si_{g,n},G^\mC)\,,~~G_V=\prod_a U^\mC_a
\eq
and
\beq{fb50}
Bun(\Si_g,G^\mC)=G_0\setminus Bun_{V}(\Si_{g,n},G^\mC)\,,
\eq
where $G_0$ is (\ref{g0}).

 We summarize the interrelations between the moduli spaces of the bundles.
Let $\clG_0$, $\clG_J$ are the gauge groups defined above  and
\beq{fgg}
G_J=\prod_a K_{a,J}\,,~~
K_{a,J}=\left\{
\begin{array}{cc}
 Id & J=0\,, \\
  C_a & J=I\,, \\
  G_a^\mR & J=II\,, \\
  U_a^\mC & J=V\,.
\end{array}
\right.
\eq
Then 
\beq{dag}
\clG_{J}(\Si_{g,n},G^\mC)=\{f\in\clG(\Si_{g},G^\mC)\,|\,f(x_a)\in G_J\}\,.
\eq
Define the finite-dimensional group
\beq{fga}
\ti G_0=\prod_aG^\mC\,.
\eq
There is the
the commutative diagram
\beq{fbj}
\xymatrix{
  \{d_{\bA}\} \ar[dr]_{\clG_J} \ar[r]^{\clG_0}
                & \quad Bun_0(\Si_{g,n},G^\mC) \ar[d]^{G_J}\ar[dr]^{\ti G_0} \\
                & Bun_J(\Si_{g,n},G^\mC) \ar[r]_{\ti G_0}& \quad Bun(\Si_{g},G^\mC)            }
\eq
Here arrows mean the passage to the quotient spaces.
The action of $\ti G_0$ is the action on the symmetric spaces $\clX_{a,J}$.
 We have a natural "forgetting" projection $Bun_J(\Si_{g,n},G^\mC)\to Bun(\Si_{g},G^\mC)$.
 The fiber of this projection is the product of the symmetric spaces $\clX_{a,J}$.

This construction is valid for
the parabolic bundles, if the gauge group $\clG_J$ is replaced by
the group $\clG_{par}$ (\ref{cpar}) and $G_J$ on
$G_{par}=\prod_aB_a^\mC$. In particular,
\beq{psm}
Bun_0(\Si_{g,n},G^\mC)\stackrel{G_{par}}{\longrightarrow}Bun_{par}(\Si_{g,n},G^\mC) \stackrel{\ti G_0}{\longrightarrow} Bun(\Si_{g},G^\mC) \,,
\eq
where again the arrows mean the passage to the quotient spaces.
On can consider arbitrary  combinations of parabolic
and quasi-antisymmetric structures. Eventually, they lead to algebraically integrable systems.
The quasi-compact (type I) and the quasi-normal (type II) systems are related to real integrable systems. We come to this point later.



\subsection{Higgs bundles}

Let $\ka$ be a canonical class on the curve $\Si_g$. The Higgs field $\Phi$ is
a $\End(E(G^\mC))\otimes\ka(\Si_g)$-valued section of the bundle of distributions on $\Si_g$.
The pair $(\Phi\,,d_{\bA})$ forms the Higgs bundle over the smooth curve $\Si_g$
\cite{Hi}. The Higgs bundle is a symplectic space with the form
\beq{40}
\Om=\int_{\Si_g}(D\Phi,D\bA)\,.
\eq
It is invariant under the transformations (\ref{fgt}) and
\beq{ht}
\Phi\to f\Phi f^{-1}\,,~~f\in\clG(\Si_g,G^\mC)\,.
\eq
The moduli spaces of the Higgs bundles are symplectic quotients with respect to this action.

Now consider the Higgs bundles over punctured curves.
 As for the moduli spaces of holomorphic bundles we have $\al$ and $\be$ realizations of these objects.


\subsubsection{ $\al$-model}

 Consider first the parabolic Higgs bundles \cite{Ko, Si}.
 To this end extend the upstairs symplectic space. The $\al$-model for the
 parabolic Higgs bundles is defined by the data
$(\Phi\,,d_{\bA},\cup \clO_a)$, where $\clO_a=\{\bfS(\nu_a,g_a)\}$ are  coadjoint $G^\mC$-orbits (\ref{orb}) attached to the marked points.
The symplectic form on this data is
\beq{sp}
\Om_{par}=\Om+\sum_a\om^{KK}_a\de(z-x_a,\bz-\bar x_a)\,,
\eq
where $\om^{KK}_a=\om^{KK}(\nu_a)$ is the Kirillov-Kostant form (\ref{kkf}).
 The corresponding Poisson brackets are
 the canonical brackets on the affine space $(D_{\bA},\Phi)$ and
  the Lie-Poisson brackets for the orbit variables.

The gauge group $\clG(\Si_g,G^\mC)$ (\ref{sgg}) is generated by  symplectomorphisms of the parabolic Higgs bundles.
It acts on the variables as
(\ref{fgt}), (\ref{ht}) and $g_a\to g_a f^{-1}(x_a)$ in the representation (\ref{kkf}) of the
Kirillov-Kostant form $\om^{KK}$.
These transformations lead to the moment map equation
\beq{mom1}
d_{\bA}\Phi+\sum_{a=1}^n\de(z-x_a,\bz-\bar x_a)\bfS(\nu_a, g_a)=0\,.
\eq
It means that in the neighborhood of the marked point $x_a$ the Higgs field has a simple pole
\beq{op}
\Phi(z)\sim \frac{\bfS_{J,a}}{z-x_a}+\bfS^0_{J,a}+(z-x_a)\bfS^1_{J,a}+\cdots\,.
\eq
Imposing the moment map constraints (\ref{mom1}) and fixing the gauge with respect
to the action (\ref{ht}) we come to the moduli space of parabolic Higgs bundles:
 \beq{msp1}
 \clM^\al_{par}(\Si_{g,n},G^\mC)=
 \left((d_{\bA}\,,\,\Phi)\,,\,\prod_{a=1}^n \clO_a\right)//\clG(\Si_g,G^\mC)
 =\{(\ref{mom1})\}/\clG(\Si_g,G^\mC)\,.
 \eq

Similarly, the Higgs bundles of types O\,,\,I\,,\,II and V are
defined by the data
\beq{3}
\begin{array}{c}
  (\Phi\,,d_{\bA});\,\prod_{a=1}^nT^*G^\mC_{a}\,,~~J=0\,, \\
  (\Phi\,,d_{\bA});\,\prod_{a=1}^nT^*\clX_{J,a}\,,~~J=I\,,\,II\,,\,V\,,
\end{array}
\eq
where $T^*\clX_{J,a}$ are the cotangent bundles to the symmetric spaces $\clX_{J,a}$.
The sections of the cotangent bundles $T^*\clX_{J,a}$ are described in terms of the variables
$g$ and $\bfX$ (\ref{xr}), or in terms of the Darboux variables $\clP$ and $\clQ$ (\ref{cav}).
These Higgs bundles are equipped with the symplectic forms
\beq{4}
\Om^J=\Om+\sum_{a=0}^n\om_a^J\de(z-x_a,\bz-\bar x_a)\,,
\eq
where $\om_a^0$ is the symplectic form on the cotangent bundle $T^*G^\mC_a$
 (\ref{sfcb}), and
$\om_a^J=\om_a^{X_J}$ is the symplectic form on the cotangent bundles $T^*\clX_{J,a}$
(\ref{sfq1}), (\ref{caf1}).

The symplectomorphisms corresponding to $\Om^J$ act as $g\rightarrow gf^{-1}$, $\bfX\rightarrow f\bfX f^{-1}$, see  (\ref{ras}).
Using (\ref{mla1}) we come to the moment map equation:
\beq{mome}
d_{\bA}\Phi+\sum_{a=1}^n\bfX_J(g_a,\zeta_a)\de(z-x_a,\bz-\bar x_a)=0\,.
\eq

The latter equation
 means that the Higgs field $\Phi$ is a meromorphic section of
the bundle $\End(E(G^\mC))\otimes\ka(\Si)$ with simple poles at $x_a$ and the residues
$\Res\,\Phi_{z=x_a}=\bfX_{J,a}$:
\beq{lh}
\Phi(z)\sim \frac{\bfX_{J,a}}{z-x_a}+\bfX^0_{J,a}+(z-x_a)\bfX^1_{J,a}+\cdots\,.
\eq
Imposing the moment constraints (\ref{mome}) and fixing the gauge  we come to the moduli space
 \beq{msp}
 \clM^\al_J(\Si_{g,n},G^\mC)=
 \left((d_{\bA}\,,\,\Phi)\,,\,\prod_{a=1}^n T^*\clX_{J,a}\right)//\clG(\Si_g,G^\mC))=
 \{(\ref{mome})\}/\clG(\Si_g,G^\mC)\,.
 \eq


\subsubsection{$\be$-model}


In the ${\bf\be}$-model the Higgs bundle is defined by the pair  $(\Phi\,,d_{\bA})$,
 the symplectic form (\ref{40}) and the gauge groups $\clG_{par}$ (\ref{cpar}),
  $\clG_{0}$ (\ref{g0}), $\clG_{I}$ (\ref{g1}), $\clG_{II}$ (\ref{g2}), $\clG_{V}$ (\ref{g5}). Then
  \beq{be1}
  \clM^\be_{par}(\Si_{g,n},G^\mC)=
(d_{\bA}\,,\,\Phi)//_\nu\clG_{par}(\Si_g,G^\mC)\,,
  \eq
  \beq{be2}
 \clM^\be_J(\Si_{g,n},G^\mC)=
 (d_{\bA}\,,\,\Phi)//\clG_{J}(\Si_g,G^\mC)\,,~~J=0,I,II,V\,.
  \eq
  Proposition A.2 claims that the $\al$-model  is equivalent to the $\be$-model.
 Let us describe this isomorphism explicitly.
Consider the one-point case.
 Let  $D_\varepsilon$ be a small disk with the local coordinate $z$ and
$z=0$ is the marked point. Then the form $\Om$ (\ref{40}) can be rewritten as
\beq{omb}
\Om=D\vartheta\,,~~\vartheta=\int_{\Si_g\setminus D_\varepsilon}(\Phi,D\bA)+\int_{ D_\varepsilon}(\Phi,D\bA)\,.
\eq
Consider the map $h=C^\infty(D_\varepsilon)\to G^\mC$.
On the disk  $D_\varepsilon$ the connection form $\bA$ can be represented as
\beq{ba}
\bA=h^{-1}\bp h\,.
\eq
  Then
 \beq{td}
 \int_{ D_\varepsilon}(\Phi,D\bA)=\int_{ D_\varepsilon}(\Phi, D(h^{-1}\bp h))\,.
 \eq
Since
$$
D(h^{-1}\bp h)=\bp(h^{-1}D h)-[h^{-1}D h,h^{-1}\bp h]\,,
$$
(\ref{td}) takes the form
$$
\int_{ D_\varepsilon}(\Phi,D\bA)=
-\int_{ D_\varepsilon}(d_{\bA}\Phi),h^{-1}D h)+
\oint_{\G_\varepsilon}(h^{-1}D h,\Phi)\,,
$$
where the contour $\G_\varepsilon$ is the boundary of $ D_\varepsilon$.
Define the gauge group
$$
 \clG_{J,\varepsilon}=\{\,f\in C^\infty(D_\varepsilon)\to G^\mC\,|\,f(0)\in K_J\}\,,
$$
where $K_J$ is (\ref{fgg}). The group $\clG_{J,\varepsilon}$ is the restriction of the gauge
group $\clG_J$ (\ref{dag}) on $D_\varepsilon$.
It acts on $\bA$ (\ref{ba}) and, therefore, on $h$ as $h\to hf^{-1}$.
Let $\ep_J\in$Lie$(\clG_{J,\varepsilon}$.
This action  is generated by the Hamiltonian
$$
\int_{\Si_g}\left(\ep_J,\bp\Phi+[\bA,\Phi]\right)+\oint_{\G_\varepsilon}(\ep_J,\Phi)\,.
$$
Let $Res_{z=0}\,\Phi(z)=\zeta_J$.    Since $\ep_J(0)\in$Lie$(K_J)=\gk_J$
the vanishing of the moment map leads to the equation
\beq{meb}
d_{\bA}\Phi+\de(z,\bz)\zeta_J=0\,,~~Pr|_{\gk^*_J}\zeta_J=0\,.
\eq
This equation coincides with the moment map constraint equation (\ref{mome})
 for $n=1$. Let in (\ref{ba}) $g=h(0)$. Then the form (\ref{omb})
 $$
 \Om=\int_{\Si_g}(D\Phi,D\bA)+D(\zeta_J,g^{-1}Dg)
 $$
coincides with (\ref{4}).


\subsubsection{Hierarchy of moduli spaces}


 The moduli space $\clM_J(\Si_{g,n},G^\mC)$ is a complex symplectic manifold
 for $J=0$ and $V$.
 Its holomorphic coordinates are $(\Phi,\bA)$, the holomorphic coordinates on  $T^*G^\mC$ are $(\zeta_0,g)$,
 and $(\zeta_V,g)$ are holomorphic coordinates on $T^*X_V$. Here  $g\in G^\mC$, $\zeta_0\in\gg^\mC$ and $Pr\,\zeta_V|_{\gu^\mC}=0$.

 From (\ref{fd}) and  (\ref{d5}) we find
 $$
 dim_\mC\,\clM_{0}(\Si_{g,n},G^\mC)=2(g-1+n)\dim\,G^\mC\,.
 $$
$$
 dim_\mC\,\clM_{V}(\Si_{g,n},G^\mC)\,=2(g-1)\dim\,G^\mC+2n\sum_{j=1}^ld_j\,.
 $$

There is the symplectic analog of diagrams (\ref{fbj}),  (\ref{psm}) for the moduli spaces $\clM_J$.
In this diagram the moduli spaces $Bun_J$ are replaced by the moduli spaces $\clM_J$ and the actions
of the group are replaced by the symplectic actions
\beq{mjm}
\clM(\Si_g,G^\mC)=\ti G_0\setminus\setminus\clM_J(\Si_{g,n},G^\mC)~~(J=0,I,II,V)\,,
\eq
\beq{mpm}
\clM(\Si_g,G^\mC)=\ti G_0\setminus\setminus\clM_{par}(\Si_{g,n},G^\mC)\,.
\eq

There exists a relation between the moduli spaces $\clM_V(\Si_{g,n},G^\mC)$ and $\clM_{par}(\Si_{g,n},G^\mC)$ based on the construction of the coadjoint orbits (\ref{oga}).
Define an analog $\clH$ of the group $G_J$ (\ref{fgg})
\beq{clh}
\clH=\prod_{a=1}^n H_a^\mC\,,
\eq
where $H_a^\mC$ are Cartan subgroups of $G^\mC$. Notice that they are simultaneously the Cartan
subgroups of the symmetric spaces  $H_a^\mC\subset\clX_{V,a}$.
We claim that
\beq{qcp}
\clM_{par}(\Si_{g,n},G^\mC)=\clH\setminus\setminus_{\vec\nu}\clM_V(\Si_{g,n},G^\mC)\,.
\eq
The action of $H_a^\mC$ on $T^*\clX_{V,a}$ is defined in (\ref{csa}).
 The  value of the moments $\vec\nu=(\nu_1,\ldots,\nu_n)$ satisfies the conditions
\beq{zej}
Pr\,|_{\gh_a^\mC}\,\zeta_{a,V}=\nu_a\,.
\eq
(see (\ref{pao})).
To prove (\ref{qcp}) consider  the coadjoint orbits $\clO_{\nu}$ (\ref{oga}) obtained by the symplectic reduction $\clO_{\nu}=H^\mC\setminus\setminus_{\nu}T^*\clX_V$.
Using the representation (\ref{msp}) for $\clM_V(\Si_{g,n},G^\mC)$ and (\ref{msp1}) for $\clM_{par}(\Si_{g,n},G^\mC)$ we derive from (\ref{oga}) the statement (\ref{qcp}).

The moduli space $\clM_0$ is a progenitor of the module spaces $\clM_J$. From the diagram (\ref{fbj}) and (\ref{psm}) we find
$$
\xymatrix{
 & \ar[dl]_{\clG_J\setminus\setminus}  \quad
  \clM_0(\Si_{g,n},G^\mC)\ar[d]^{\clG_{par}\setminus\setminus}
                \ar[dr]^{G_0\setminus\setminus}& \\
   \clM_J(\Si_{g,n},G^\mC)& \clM_{par}(\Si_{g,n},G^\mC) & \quad \clM(\Si_{g},G^\mC)
   }
$$
where the arrows mean the symplectic quotients. As will be shown below,
 $\clM_{I,II}$ eventually lead to real integrable systems.

 The insertion of cotangent bundles $T^*X_{J,a}$ for $J=I\,,\,II$ at the points $x_1,\dots,x_q$
  breaks the complex structures of the Higgs bundles.
 The moduli spaces  $\clM_{I,II}(\Si_{g,n},G^\mC)$ are
   real symplectic spaces.  Their  dimensions (see (\ref{dI}), (\ref{dII})) are
 \beq{dhb}
\dim_\mR\,\clM_{I}(\Si_{g,n},G^\mC)=\dim_\mR\,\clM_{II}(\Si_{g,n},G^\mC)=
 \eq
$$
=2(g-1)\dim_\mR\,G^\mC+2n\dim_\mR \,\clX_{I,II}=2(g-1)\dim_\mR\,G^\mC
+2n\sum_{j=1}^l(2d_j-1)\,.
$$


\subsection{Integrals of motion}

 Let ${\cal V}$ be the algebra of vector fields on $\Si_{g,n}$ vanishing at the marked points.
Consider the generalized Beltrami differentials as elements of the cohomology group
\beq{gbd}
\mu_{d_j}\in H^1(\Si_{g,n},{\cal V}^{\otimes (d_j-1)}\otimes\bar \ka)\,,
\eq
where $\bar \ka$ is the anti-canonical class
on $\Si_{g,n}$ and $d_j$ are the orders of invariants of the algebra $\gg^\mC$.
In particular, $\mu_2$ is the standard Beltrami differential.
Let
$n_j=\dim\, H^1(\Si_{g,n},{\cal V}^{\otimes (d_j-1)}\otimes\bar \ka)$
and  $\mu^k_{d_j}$, $k=0,\ldots, d_j-1$ be a basis in $H^1(\Si_g,{\cal V}^{\otimes (d_j-1)}\otimes\bar \ka)$.
From the Riemann-Roch theorem
\beq{nj}
n_j=(2d_j-1)(g-1)+n d_j\,,
\eq
where the last term is the contribution of the marked points.
By means of the generalized Beltrami differentials  we construct the gauge invariant quantities
\beq{5}
I_{jk}=\int_{\Si_g}(\Phi^{d_j})\mu^k_{d_j}\,.
\eq
They are independent and Poisson commute with respect to brackets coming from the
symplectic form (\ref{4}). In accordance with the Liouville theorem for complete
integrability we need $\dim\,(Bun)$ integrals.
From (\ref{nj}) the total number of integrals is
$$
\clN_G=\sum_{j=1}^ln_j=(g-1)\sum_{j=1}^l(2d_j-1)+n\sum_{j=1}^ld_j
$$
or (see (\ref{dcg}), (\ref{df}))
\beq{ni}
\clN_G=(g-1)\dim_\mC\,G^\mC+n\sum_{j=1}^ld_j=(g-1)\dim_\mC\,G^\mC+n(\dim_\mC\,(Fl(G^\mC))+l)\,.
\eq
It seems that for the parabolic bundles we have $nl$ extra integrals of motion
(see (\ref{fd1})).
This is actually not the case. To demonstrate it consider
 the dependence of differentials $\mu$ on  positions of
the marked points. Let ${\mathcal U}'_a$ be
 neighborhoods
 of the marked points $x_a\,$, $(a=1,\ldots,n)$
such that ${\mathcal   U}'_a\cap{\mathcal   U}'_b=\emptyset$ for $a\neq b$.
Define a smooth function $\chi_a(z,\bz)$
\beq{cf}
\chi_a(z,\bz)=\left\{
\begin{array}{cl}
1,&\mbox{$z\in{\mathcal   U}_a$ },~{\mathcal   U}'_a\supset{\mathcal   U}_a\\
0,&\mbox{$z\in\Si_{g,n}\setminus {\mathcal   U}'_a$}\,.
\end{array}
\right.
\end{equation}
In a neighborhood of the marked point $x_a$ the differentials behave as
\beq{mua}
\mu_{d_j,a}=\sum_{k=0}^{d_j-1}\mu^{d_j}_{k,a}\,,~~\,\mu^{d_j}_{k,a}=
t^{d_j}_{a,k}(z-x_a)^k\bp\chi_a(z,\bz)\,.
\eq
Here $t^{d_j}_{a,k}$ can be considered as coordinates on the space
$H^1(\Si_g,{\cal V}^{\otimes (d_j-1)}\otimes\bar \ka)$.
The corresponding integrals (\ref{5}) assume the form:
\beq{imp}
I_{j,k,a}\sim\int_{\Si_g}(\Phi^{d_j})\mu^{ d_j}_{k,a}
\sim\int_{\Si_g}(\Phi^{d_j})(z-x_a)^k\bp\chi_a(z,\bz)\,.
\eq
Then from (\ref{lh}) we conclude
\beq{inte}
I_{j,d_j-1,a}\sim (\bfT_a)^{d_j}\,,~~
I_{j,d_j-2,a}\sim\sum_{b\neq a}\frac{(\bfT_a^{d_j-1}\bfT_b)}{x_a-x_b}+
\sum_b(\bfT_a^{d_j-1}\bfT^0_b)+\cdots\,,
\eq
where $\bfT_a=\bfS_a$ for the parabolic bundles and $\bfT_a=\bfX_{J,a}$ for the bundles of type $J$.

For the parabolic bundles
 the integrals $I_{j,d_j-1,a}\sim(\bfS_a^{d_j})$ $(j=1,\ldots,l)$, $(a=1,\ldots,n)$
are the Casimir functions defining the orbits and thereby they are irrelevant.
In this way we come to the Liouville integrability in the parabolic case.

For the bundles of type V there are the needed number of integrals of motion, that is $\clN_G=\dim\,Bun_V(\Si_{g,n},G^\mC)$ (see (\ref{bV})).
For these bundles $I_{j,d_j-1,a}\sim(\bfX_{V,a}^{d_j})$ are no longer the  Casimir functions.
Thus, the moduli space of Higgs quasi-antisymmetric bundles is a phase space of integrable
systems.

It follows from this construction that there exist the mixed complex integrable systems with insertion
of coadjoint orbits $\clO_a$ at some marked points and the cotangent bundles $T^*\clX_{V,b}$
at other marked points.

As we noticed, the moduli space of the Higgs bundles of type I and II are real spaces.
The integrals related to the marked points (\ref{imp}) are real quantities, because the quantities
$(\bfT_a)^{d_j}$, $(\bfT_a^{d_j-1}\bfT_b),\ldots$ are real.
We should compare
\beq{tng}
\clN^\mR_G=(g-1)\dim_\mR\,G^\mC+n\sum_{j=1}^ld_j
\eq
and (\ref{dI}), (\ref{dII})
$$
\dim_\mR\,Bun_{I,II}(\Si_{g,n},G^\mC)=(g-1)\dim_\mR\,G^\mC+n \sum_{j=1}^l( 2d_j-1)\,.
$$
Then we have the deficiency $\de_{G}$ of integrals
for the system to be completely integrable:
\beq{def}
\de_{G}=\dim_\mR\,Bun_{I,II}(\Si_{g,n},G^\mC)-\clN^\mR_G=n\sum_{j=1}^l(d_j-1)\,.
\eq
In the next Section we come to integrable systems by passing to the real forms of types III and IV.



\section{Real structures}

\setcounter{equation}{0}

Assume that  the Riemann surface $\Si_g$ is equipped with an anti-holomorphic
involution
$$\jmath\,:\,\Si_g\to\Si_g.$$  In a neighborhood of a fixed point
one can find a local
coordinate $z$ such that the involution $\iota$ is given by the complex conjugation
$\jmath(z)=\bz$ \cite{GH}. The fixed point set can be seen as a union
$\Si^0_g=\cup S^1_j$ of copies of the unit circle $S^1$ embedded in $\Si_g$.
We assume that the set of marked points  $\clD=(x_1,\ldots,x_n)$ is invariant with respect to the involution
\beq{id}
\jmath(\clD)=\clD\,.
\eq
There are two possibilities:
\beq{cdd}
1.\,\jmath(x_a)=x_a\,,~(a=1,\ldots,n)\,,\qquad
2.\,\jmath(x_a)=x_{a+1}\,,~x_a\,,\,x_{a+1}\in\clD\,.
\eq
Either $\clD\subset \Si^0_g$  or some elements of $\clD$
are conjugated under the involution $\iota$.

\subsection{From Type I to Type IV}

Let $d\si$ be the antiholomorphic involutive automorphism (\ref{si}) acting on
the complex Lie algebra $\gg^\mC$. The fixed point set of $\gg^\mC$
 is the normal real form of $\gg^\mR$.
 Attach to the points from $\clD$ the cotangent bundles $T^*\clX_{I,a}$.
 Following the general scheme \cite{BS} we combine the anti-holomorphic involution of the curve
 $\jmath$
with an involutive automorphism $\si$ of the group $G^\mC$ into the operator $i_\Si$ acting 
on the Higgs data of the moduli space $\clM_{I}(\Si^0_{g,n},G^\mC)$
\footnote{In \cite{BS} it is denoted as $i_3$.}:
\beq{01}
\imath^\si(\Phi\,,d_{\bA}(z)\,,\,\prod T^*\clX_{I,a})=
(\jmath^*\si(\Phi)\,,\jmath^*\si(d_{\bA})\,,\,\prod T^*\si(\clX_{I,a}))\,.
\eq

First, consider the case 1 in (\ref{cdd}) ($\clD\subset \Si^0_g$).
Since the invariant subgroups of $G^\mC$ and $C_a$ with respect to $\si$ are $G^\mR$ and $U_a$,
the fixed point set of the $\si$-action at the marked points is the product of cotangent bundles
to the symmetric spaces $\prod (T^*\clX_{IV,a}=T^*(U_a\setminus G^\mR)$.
%
In this way we come to the moduli space $\clM_{IV}(\Si^0_{g,n},G^\mR)$ of $G^\mR$-Higgs bundles
over $\Si^0_{g,n}$ with the cotangent bundles $T^*\clX_{IV,a}$ attached to the marked points.
Its real dimension equals
\beq{fpbd}
\dim_\mR\,\clM_{IV}(\Si^0_{g,n},G^\mR)=
2(g-1)\dim
\,G^\mR+2n\dim \,\clX_{IV} \,.
\eq
Let $\Phi^0$ be the invariant Higgs field $\Phi^0=\iota^*(\si(\Phi^0))$.
The involution $\imath^\si$ acts on the generalized Beltrami differentials (\ref{gbd}) and,
in particular, on the differentials related to the marked points (\ref{mua})
\beq{bdi}
\mu\to\jmath^*(\mu)\,.
\eq
Let $\mu(0)^k_{d_j}$ be the  differentials, which are invariant with respect
 to this involution. Then the quantities  (\ref{5})
\beq{5b}
I^0_{jk}=\int_{\Si_g}((\Phi^0)^{d_j})\mu(0)^k_{d_j}
\eq
 are independent involutive real integrals of motion.
 Since $\dim_\mR\,G^\mR=\dim_\mC\,G^\mC$\,
for the number of integrals we have
\beq{ng0}
\clN^\mR_G=
(g-1)\dim_\mR\,G^\mR+n\sum_{j=1}^ld_j\,.
\eq
From (\ref{dgxr}) and (\ref{fpbd}) we
find that
\beq{d4}
\oh\dim_\mR\,\clM_{IV}(\Si^0_{g,n},G^\mR)=\clN_G^\mR\,.
\eq
It means that the moduli space $\clM_{IV}$ is the phase space of a
real integrable system with the hierarchy of commuting  integrals (\ref{5b}).

Now consider the case 2. in (\ref{cdd}). For simplicity take $n=2$,
Locally the fixed point set is  $\Im m\,z=0$ and $x_2=\bar x_1=\bar x$.
We have two cotangent bundles $T^*\clX_{I,1}$ and $T^*\clX_{I,2}$ attached to these points,
where $\clX_{I,2}=\si(\clX_{I,1})$. In total we have $\sum_j(2d_j-1)$ degrees of freedom (\ref{dgx}).
Using representation (\ref{lh}) we write the Higgs field as
\beq{n2h}
\Phi(z)\sim \frac{\bfX^{(-1)}_{I}}{z-x}+\bfX^{(0)}_{I}+(z-x)\bfX^{(1)}_{I}+
\frac{\si(\bfX^{(-1)}_{I})}{z-\bar x}+\si(\bfX^{(0)}_{I})+(z-\bar x)\si(\bfX^{(1)}_{I})+\cdots\,.
\eq
As in (\ref{inte}) we have two families of integrals related to the points $x=x_1$ and $x_2=\bar x$
$$
I_{j,d_j-1,1}\sim (\bfX^{(-1)}_I)^{d_j}\,,~~
I_{j,d_j-2,1}\sim
\frac{((\bfX^{(-1)}_I)^{d_j-1}\si(\bfX_I))}{x-\bar x}+
((\bfX^{(-1)}_I)^{d_j-1}(\si(\bfX^{(0)}_I)+\bfX^{(0)}_I))+\cdots\,,
$$
$$
I_{j,d_j-1,2}\sim (\si\bfX^{(-1)}_I)^{d_j}\,,~~
I_{j,d_j-2,2}\sim
\frac{((\si(\bfX^{(-1)}_I))^{d_j-1}\bfX_I)}{\bar x-x}+((\si(\bfX^{(-1)}_I))^{d_j-1}(\bfX^{(0)}_I)+
\si(\bfX^{(0)}_I))\cdots\,.
$$
The integrals
$I_{j,d_j-1,1}=(\bfX^{(-1)}_I)^{d_j}$ and $I_{j,d_j-1,2}=(\si\bfX^{(-1)}_I)^{d_j}$ coincide.
To justify this fact we use representation (\ref{xr})  $\bfX_J=g\zeta_I g^{-1}$.
There is a basis in the Lie algebra $\gg^\mC$ such that $\zeta_I$ are  traceless Hermitian
matrices and $\si$ acts as the transposition. In this way these integrals are equal.
Thereby, we have $\sum_j(2d_j-1)$ integrals. They provide the complete integrability, because
this number is equal to the dimension of $\clX_I$ (\ref{dgx}).

\subsection{From Type II to Type III}

Consider the involutive automorphism $d\rho$ of the algebra $\gg^\mC$, which was the fixed set subalgebra. It
is the compact form $\gc$ (\ref{rho}).
Similarly to (\ref{01}) define the operator $\imath^\rho$ acting on the moduli space
$\clM_{II}(\Si^0_{g,n},G^\mC)$:
\beq{010}
\imath^\rho(\Phi\,,d_{\bA}(z)\,,\,\prod T^*\clX_{II,a})=
(\jmath^*\rho(\Phi)\,,\jmath^*\rho(d_{\bA})\,,\,\prod T^*\rho(\clX_{II,a}))\,.
\eq
In the case 1. (\ref{cdd}) the fixed points set of the $\imath^\rho$ action is the moduli space
$\clM_{III}(\Si^0_{g,n},C)$ of
the Higgs $C$-bundles over $\Si^0_{g,n}$ with the cotangent bundles $T^*\clX_{III,a}$
($\clX_{III,a}=U_a\setminus C$) attached to the marked points. The symmetric spaces
$\clX_{III,a}=U_a\setminus C$  and
$\clX_{IV,a}=U_a\setminus G^\mR$ are Cartan dual and have the same dimensions.
Since the $\imath^\rho$ action on the generalized Beltrami differentials still have the form (\ref{bdi})
we have (as above) the same number of invariant integrals.
Therefore,  as in
(\ref{d4}) we have
$$
\oh\dim_\mR\,\clM_{III}(\Si^0_{g,n},C)=\clN_G^\mR\
$$
and we come to the real completely integrable systems of type III.

In the case 2 (\ref{cdd}) we have the same form of the Higgs field as in (\ref{n2h}), where
$\bfX_I$ is replaced by $\bfX_{II}$ and the involutive automorphism $\si$ -- by the involutive automorphism $\rho$. To find the correct number of integrals of motion
one should prove that
$(\bfX^{(-1)}_{II})^{d_j}=(\rho(\bfX^{(-1)}_{II}))^{d_j}$. Again we use representation $\bfX_{II}=g\zeta_{II}g^{-1}$, where $\zeta_{II}$ can be represented as traceless imaginary matrices and $\rho$ acts as a transposition. In this way we come to the needed equality, and thereby, to the complete
integrability for this structure of poles of the Higgs field.

\bigskip
The complexification of the moduli spaces of types III and IV is the moduli space $\clM_V$
(\ref{msp}). Acting in the opposite direction one can start with the moduli space $\clM_V$, and then by applying the involutions
$\imath^\rho$ and $\imath^\si$ obtain (as the fixed point sets) real moduli spaces $\clM_{III}$ and
$\clM_{IV}$. Taking into account the diagram (\ref{ssr}) we describe the real reductions by the diagram:
\beq{msr}
\xymatrix{
  \clM_{II} &
   & \clM_V  & & \clM_I          \\
 &\ar[ul]_{\imath^\rho}\clM_{III}\ar[ur]^{\imath^\rho} & &\ar[ul]_{\imath^\si}     \clM_{IV} \ar[ur]^{\imath^\si}        }
 \eq
where the arrows mean the embeddings as fixed  points sets of the involutions.


\subsection{Mixed systems}

In the complex case we have constructed the mixed integrable systems of parabolic and
quasi-antisymmetric types. Here we
 consider the mixed systems of types III or IV with the systems with parabolic
singularities.  Assume that corresponding marked point are located on the invariant
subset $\Si^0_{g,n}$.

Consider first the action of the involution $\imath^\si$ on the parabolic bundles.
The invariant subgroup of this action is the normal form  $G^\mR$, and we
come to the $G^\mR$-bundles over  $\Si^0_{g,n}$. It means that at the marked points
we have $B^\mR$ flag varieties $\Fl_a^\mR=B^\mR\setminus G^\mR$, where $B^\mR=B^\mC\cap G^\mR$.
For the Higgs bundles they correspond to the $G^\mR$ coadjoint orbits $\clO^\mR_a$.
Thus we  come to the fixed point set $\clM_{par}(\Si^0_{g,n},G^\mR)$. It is a phase space of real
integrable systems. It means that there exist
 mixed systems  $\clM_{par,IV}(\Si^0_{g,n},G^\mR)$
  with insertion of real orbits  $\clO^\mR_a$ at some marked points and cotangent bundles
$T^*\clX_{IV,b}$ at the remaining points.

The action of the involution $\imath^\rho$ leads to the $C$-bundles over  $\Si^0_{g,n}$.
Then the intersection of the complex orbits with the compact subalgebra $\clO_a\cap \gc$
is isomorphic to the orbits of compact subgroup $\clO^C_a=T_a\setminus C$, where $T_a=B^\mC\cap C$ is the Cartan torus
of the compact group $C$.  Due to the
Iwasawa decomposition $G^\mC=CB^\mC$ the latter is isomorphic to a complex flag variety
 $Fl_a^\mC=B^\mC\setminus G^\mC$.
Therefore, the fixed point set is the moduli space $\clM_{par}(\Si^0_{g,n},C)$ of the Higgs
$C$-bundles over $\Si^0_{g,n}$ with the  flag varieties $Fl_a^\mC$   attached to some marked points
and cotangent bundles
$T^*\clX_{III,b}$ at the remaining points.






\section{Double coset construction}

\setcounter{equation}{0}

Let us give an
alternative descriptions of the moduli spaces
$Bun_{J}(\Si_{g,n},G^\mC)$ and $\clM_{J}(\Si_{g,n},G^\mC)$, $J=0,\ldots V$.
It is based on the double coset construction \cite{BD02,DS} of
$Bun_{par}(\Si_{g,n},G^\mC)$ and $\clM_{par}(\Si_{g,n},G^\mC)$.
For simplicity we assume that
 $G^\mC$ is  simply-connected, although in fact everything works in the general case.

 As above, there are two equivalent ways to define these
moduli spaces. We describe both of them (models $\al$ and $\be$). Before coming to the moduli spaces we present
two constructions of the affine flag varieties and affine symmetric spaces.

\subsection{Affine flags and affine symmetric spaces}

First consider the $\be$-model description of the affine spaces (Appendix A).
Let $K$ be a subgroup of $G^\mC$. We will consider two cases. First $K=B^\mC$ for the flag variety
 $Fl^\mC=B^\mC\setminus G^\mC$. and the second $K=K_J$
\beq{sk}
K_0=Id\,,\, K_I=C\,,\,K_{II}=G^\mR\,,\,K_V=U^\mC
\eq
(see Table 1). The quotients
 $\clX_J=K_J\setminus G^\mC$ are the symmetric spaces from Table 1.
In fact, the following construction  does not depend on the choice of $K_J$.
Define the loop group
$$
L(G^\mC)=G^\mC\otimes\mC[t^{-1},t]]\,,
$$
 its subgroups
$$
L^+(G^\mC)=G^\mC\otimes\mC[[t]]\,,
$$
\beq{lt}
L^+(G^\mC,K)=\{K(Id+t L^+(G^\mC))\}
\eq
and the quotient space $\clK^{aff}(K)=L(G^\mC)/L^+(G^\mC,K)$.
In particular,
\begin{subequations}\label{y}
  \begin{align}
& {\rm Affine~flag~ variety}~ \Fl^{aff}=L(G^\mC)/L^+(G^\mC,B^\mC)\,,\\
&{\rm Affine~symmetric~ space} ~\clX_J^{aff}(K_J)=L(G^\mC)/L^+(G^\mC,K_J)\,,~~J=I,II,V\,.
 \end{align}
\end{subequations}
\begin{rem}
The affine flag variety is a particular case of the affine Grassmannian \cite{PS}.
The definition of the affine symmetric space $\clK^{aff}(K_J)$ differs from the conventional one \cite{Hei}.
\end{rem}

Let $P^\vee\subset\gh^\mC$ be the coweight lattice (see Appendix B).
Consider the affine Weyl group  $$W^{aff}=\{\hat w=wt^\ga\,|\,w\in W\,,\, \ga\in P^\vee\}$$ and
define the subgroup $L^-(G^\mC)\subset L(G^\mC)$
$$
L^-(G^\mC)=\{n_-(Id+t^{-1}G^\mC\otimes[t^{-1}])\,,~n_-\in N^-\}\,.
$$
The affine Bruhat decomposition \cite{PS}
$$
L(G^\mC)=\left(L^-(G^\mC)\bigcup_{\hat w\in W^{aff}}\hat w\right) L^+(G^\mC,B^\mC)
$$
is the generalization of the Bruhat decomposition (\ref{GD}). Then we have explicit description
of the  Affine flag variety (\ref{y}a)
$$
 \Fl^{aff}=  \left(L^-(G^\mC)\bigcup_{\hat w\in W^{aff}}\hat w\right)\,.
$$
This representation defines the stratification of $ \Fl^{aff}$ on the Schubert cells corresponding to the elements of the affine Weyl group.




An alternative definition of these quotient spaces is based on  the $\al$-model.
We identify
$$
X=L(G^\mC)\,,~~  H=L^+(G^\mC,K)\,,~~G=L^+(G^\mC)\,.
$$
Then the $\al$-model is the quotient $(L(G^\mC)\times(L^+(G^\mC,K)\setminus L^+(G^\mC)))/L^+(G^\mC)$.
 Its  equivalence to the $\be$-model is provided by the Proposition A.1.
In these terms the Proposition claims that
$$
\clK^{aff}(K)=\left(L(G^\mC)\times\left(L^+(G^\mC,K)\setminus L^+(G^\mC)\right)\right)/L^+(G^\mC)\,.
$$
Notice that
$$
L^+(G^\mC,K)\setminus L^+(G^\mC)\sim K\setminus G^\mC=\clX~ {\rm or}~ \Fl^\mC\,.
$$
Thereby (compare with (\ref{y}))
\begin{subequations}\label{cl2}
  \begin{align}
&\Fl^{aff}=\left(L(G^\mC)\times\Fl^\mC\right)/L^+(G^\mC)\,,\\
&\clX_J^{aff}=\left(L(G^\mC)\times \clX_J\right)/L^+(G^\mC)\,.
\end{align}
\end{subequations}


\subsection{Moduli space  $\Bun(\Si_{g,n},G^\mC)$}

Let $\clD=(x_1,\ldots,x_n)$ be the set of marked points.
Using the $\be$-model description define
the moduli space  $\Bun(\Si_{g,n},G^\mC)$ in the following way.

The $E(G^\mC)$-bundle can be trivialized over small disjoint disks $D=\cup_{a=1}^nD_a$ around the
marked points and over the complement to the set of the marked points $\Si_{g,n}\setminus \clD$.
Then $E(G^\mC)$ bundle is defined by the transition holomorphic functions on
 $$
 D^\times=\cup_{a=1}^n(D_a^\times)\,,~~
 D_a^\times=D_a\setminus x_a\,.
 $$
  Let $G^\mC(X)$ be the holomorphic maps from $X\subset\Si_{g,n}$ to $G^\mC$.
  Describe these maps in terms of  local coordinates  $t_a$ on  the disks $D_a$.
The  group $G^\mC(D)$ has the form of the $G^\mC$-valued polynomials
 \beq{lre}
G^\mC(D)=\prod_{a=1}^nG^\mC(D_a)=\prod_{a=1}^nL_a^+(G^\mC)\,,~~L_a^+(G^\mC)=G^\mC\otimes\mC[[t_a]]\,.
 \eq
 To simplify notations in the definition of
 $G^\mC(D^\times)$ we consider first the one point case. Let $\ga$ be an element of the coweight lattice $P^\vee$ .
Define the group of maps
 \beq{ma}
 G_\ga^\mC(D^\times)=\{k_\ga(t)\in t^{\ga}L(G^\mC)\,|\,L(G^\mC)=G^\mC\otimes\mC[t^{-1},t]]\} \,.
  \eq
  If $\ga\in\clQ^\vee$ then $k(t)$ has a trivial monodromy $k(t\exp\,2\pi\imath)=k(t)$.
  Otherwise, the monodromy is nontrivial
   $k(t\exp\,2\pi\imath)=\exp\,(\frac{2\pi\imath}m)k(t)$, where $\exp\,(\frac{2\pi\imath}m)$
   represents an element of the center $\clZ(G^\mC)\sim \clP^\vee/\clQ^\vee$.
 It means that $k(t)$ is the map of $(D^\times)$ to the adjoint group $G^{ad}$
  (\ref{adj}), but
  not to $G^\mC$. In this case there is an obstruction  described by the second cohomology group $H^2(\Si,\clZ(G^\mC))\sim\clZ(G^\mC)$
   to lift the $G^{ad}$-bundle  to $G^\mC$-bundle.
   The group $H^2(\Si,\clZ(G^\mC))$ defines \emph{the characteristic classes} of the holomorphic bundles
   \cite{Ra}. The construction of the corresponding Higgs bundles and related Hitchin systems were
  discussed in \cite{LOZ1}.
  In the general case
   \beq{ma1}
 G_{\vec\ga}^\mC(D^\times)=
 \prod_{a=1}^nt_a^{\ga_a}L_a(G^\mC)=\prod_{a=1}^n\{k_{\ga_a}(t_a)\}\,,~~\vec\ga=(\ga_1,\ldots,\ga_n) \,.
  \eq
  The characteristic class of the bundle is defined by the sum $\sum_a\ga_a$.


Define two gauge groups
  \beq{gout}
\clG_{out}=G^\mC(\Si_{g}\setminus \clD)
\eq
 and
  \beq{gin}
\clG_{int}=G^\mC(D)=\prod_{a=1}^n L^+_a(G^\mC)\,,~~ L^+_a(G^\mC)=
G^\mC\otimes\mC[[t_a]]\,,~t_a\in D_a\,.
\eq
 Modify the last group as
 \beq{gt}
 \widetilde\clG^J_{int}=\prod_{a=1}^n  L^+_a(G^\mC,K_{a})
 \,,~~(\widetilde\clG^J_{int}\subset\clG_{int})\,,
 \eq
where $L^+_a(G^\mC,K_{a})$ is defined through (\ref{lt}), and $K_a$ is either equal to $B_a^\mC$, or has the form (\ref{sk})
$K_a=K_{J,a}$.

The group $\clG_{out}$ is holomorphic on  $\Si_{g}\setminus \clD$
and acts from left on $G^\mC(D^\times)$. The group $\clG^J_{int}$ acts from
the right.
Define the moduli spaces in the $\be$-model description
\beq{dcc}
 \Bun_{J,\ga}(\Si_g\setminus\clD,G^\mC)= \clG_{out}  \setminus \left(\prod_{a=1}^n t_a^{\ga_a}L_a(G^\mC) /L^+_a(G^\mC,K_{a}) \right)\,.
   \eq
  For $\ga_a\in Q^\vee$ this definition is equivalent to the given above for $Bun_{par}$
(\ref{mpar}) and  $\Bun_J$
(\ref{t00}), (\ref{bI}), (\ref{bII}), (\ref{bV}). In this case the connection $d_{\bA}$ can be reconstructed  from $k(t)$.

Due to (\ref{y}) and (\ref{gin}) the moduli spaces for the one marked point case can be rewritten as
in the $\al$-model:
\beq{af1}
\begin{array}{l}
   \Bun_{par,\ga}(\Si_{g,n},G^\mC)=\clG_{out}  \setminus
 \left(t^\ga(\Fl^\mC)^{aff}\right)\equiv\clG_{out}  \setminus
 \left((t^\ga L(G^\mC)\times\Fl^\mC)\right)/\clG_{int}\,, \\
   \Bun_{J}(\Si_{g,n},G^\mC)=\clG_{out}  \setminus
 \left(t^\ga\clX_{J}^{aff}\right)\equiv\clG_{out}  \setminus
 t^\ga\left(L(G^\mC)\times\clX_{J}\right)/\clG_{int}\,.
\end{array}
\eq



\subsection{Moduli space  of Higgs bundles  $\clM(\Si_{g,n},G^\mC)$}

To define the moduli space of the Higgs bundles  $\clM_{par,\ga}(\Si_{g,n},G^\mC)$ and $\clM_{J,\ga}(\Si_{g,n},G^\mC)$ in the double coset construction
we introduce analogues of the Higgs fields
\beq{hf}
\eta_a(t_a)\in (L_a(\gg^\mC)\otimes dt_a)\,, ~~a=1,\ldots,n\,.
\eq
and replace the symmetric spaces and the flag varieties with the corresponding cotangent bundles and
the coadjoint orbits respectively. In the $\al$-model the Higgs data corresponding to the description of
(\ref{af1}) are of the form:
$$
\bigcup_a\,\left(\eta_a,t^{\ga_a}k_a(t_a))\,,\,\left(
\begin{array}{ll}
  T^*\clX_{J,a} & J=0,I,II,V\,, \\
  \clO_a &
\end{array}
\right)\right)\,,~k_a(t_a)\in  L_a(G^\mC)\,.
$$
These data are endowed with the symplectic form
\beq{omj}
\Om_{J}=\sum_a\left(\oint_{\G_a} D(\eta_ak_a^{-1}Dk_a)+\om_a\right)\,.
\eq
Here $\G_a$ is a contour around $t_a=0$ in $D^\times_a$, $\om_a=\om^{KK}$ (\ref{kkf}) for
the parabolic bundles, and $\om_a=\ti\om(\clX_{J,a})$ (\ref{sfcb}) or (\ref{sfq1}) for the cotangent bundles.

The action of
$ \clG_{int}=\prod_aG^\mC(D_a)$ (\ref{gin}) and $\clG_{out}=G^\mC(\Si_g\setminus\clD)$
$G^\mC(D^\times)$   is lifted to the Higgs fields (\ref{hf}) as
 \beq{gah}
\clG_{a,int}\,:\,\eta_a\to f_{a,int}\eta_af^{-1}_{a,int}\,,~~k_a\to k_af^{-1}_{a,int}\,,
\eq
$$
\bfX_{J,a}\to f_{a,int}(0)\bfX_{J,a}f_{a,int}^{-1}(0)\,,~~
\bfS_a\to   f_{a,int}(0)\bfS_a f^{-1}_{a,int}(0) \,,~~g_a\to f_{a,int}(0)g_a\,,
$$
 \beq{gah1}
\clG_{out}\,:\,\eta_a\to \eta_a\,,~k_a\to f_{a,out}k_a\,.
 \eq
The $\clG_{out}$-action is generated by the Hamiltonian
\beq{hout}
F_{out}=\sum_a\oint_{\ga_a}(\ep_{out},\Ad_{k^{-1}_a}\eta_a)\,,
 \eq
where $\ep_{out}\in Lie(G(\Si_{g}\setminus \clD)$. The corresponding  moment map is
 \beq{muo}
\mu_{out}=Pr_{Lie^*(G(\Si_{g}\setminus \clD)})\sum_a\Ad_{k^{-1}_a}\eta_a(t)\,.
 \eq
The moment map constraint $\mu_{out}=0$ means that
\beq{mc1}
\Ad_{k^{-1}_a}\eta_a~{\rm has~ holomorphic ~continuation~ on~}
\Si_{g}\setminus \clD\,.
\eq
The action of $\clG_{a,int}$ on $\om_a$ leads to the moments $\mu^L$ (\ref{lm2}) and  (\ref{mla1}).
  Then we come to the Hamiltonians
 \beq{hint}
F_{a,int}=\left\{
\begin{array}{ll}
  \oint_{\G_a}(\ep_{a,int},\eta_a-\bfX_{J,a})\,,& {\rm for~} \om_a=\ti\om_a\,,\\
   \oint_{\G_a}(\ep_{a,int},\eta_a-\bfS_a)\,, & {\rm for~} \om_a=\om_a^{KK}\,,
\end{array}
\right.
\eq
 where $\ep_{a,int}\in Lie(L_a^+(G\mC))$.


Similarly the condition $\mu_{int}=0$ implies that $\eta_a$ can be continued on $D_a$ with simple poles at the marked points $t_a=0$ and
\beq{res}
Res\,\eta_a|_{t_a=0}=\left\{
\begin{array}{ll}
   \bfX_{J,a} & J=0,I,II,V\,, \\
 \bfS_a & \,.
\end{array}
\right.
\eq
The conditions (\ref{mc1}), (\ref{res}) define the moduli space as the symplectic quotient.
For a single marked point case we have
 $$
\clM_{J,\ga}(\Si_g\setminus\clD,G^\mC)=
\clG_{out} \setminus\setminus(\eta,t^\ga T^*\bfX^{aff}_{J}\,,\,\bfS)//\clG_{int}\,,
~~k(t)\in t^\ga L(G^\mC)\,.
 $$
 Similarly to (\ref{af1}) we have
 \beq{dcm}
 \begin{array}{l}
   \clM_{par,\vec\ga}(\Si_{g,n},G^\mC)=\clG_{out}  \setminus
 \left(\prod_{a}(t_a^{\ga_a},\clO^\mC_a)^{aff})\right)\,, \\
   \clM_{J,\vec\ga}(\Si_{g,n},G^\mC)=\clG_{out}  \setminus
 \left(\prod_{a}(t_a^{\ga_a},T^*\clX_{J,a}^{aff})\right)\,.
\end{array}
\eq
It was proved in \cite{LO} that
the data which were used to define the moduli spaces $\clM^\al_{par}(\Si_{g,n},G^\mC)$ in the previous description (\ref{msp1}) can
be reconstructed from $\clM_{par,\vec\ga}(\Si_{g,n},G^\mC)$.

\bigskip

Consider the $\be$-model construction. We replace
the form $\Om_j$ (\ref{omj}) on
$$
\ti\Om_{J}=\sum_a\oint_{\G_a} D(\eta_ak_a^{-1}Dk_a)\,,
$$
the gauge group $\clG^J_{int}$ (\ref{gin}) on
 $\widetilde\clG^J_{int}$ (\ref{gt}).
Similarly to (\ref{hint}) the Hamiltonian generating the gauge transformation is of the form:
\beq{hint1}
\ti F_{a,int}=
  \oint_{\G_a}(\ti\ep_{a,int},\eta_a)\,,
\eq
where
\beq{lin}
\ti\ep_{a,int}\in Lie(L_a^+(G^\mC))\,,~~\ti\ep_{a,int}=x_{a,0}+t_ax_{a,1}+\ldots\,.
 \eq
  $$
 x_{a,0}\in\,Lie\,(B^\mC)=\gb\,,~{\rm or~}Lie\,(K_{J,a})=\gk_J\,.
 $$
The moment map constraints $\ti F_{a,int}=0$ assume the form
\beq{be11}
1.\,Pr_{\gk^*_J}\eta_a=0\,,~{\rm or}~ 2.\,Pr_{\gb^*_J}\eta_a=\nu\in\gh^\mC\,.
\eq
In the case 1. the symplectic quotient is described by the pairs $(\eta_a,k_a)$, where
$\eta_a$ satisfies this condition, and due to (\ref{gah}) $k_a$ is invariant under the action
$k_a\to k_af^{-1}_a$, $f_a\in K_{J,a}$.
This pair defines  the cotangent $T^*\clX_{J,a}$ as in (\ref{mcc}), (\ref{pv}), where the right action
of the gauge group is replaced on the left one.

In the case 2. we come to the coadjoint orbit (\ref{bo}) passing through the Cartan element
$\nu\in\gh^\mC$.


\subsection{Real structures}

Here we investigate the passages from the moduli spaces $\clM_{I,II}$ to $\clM_{III,IV}$ in terms
of the double coset construction. In section 3 it was performed for  trivial bundles.
We  investigate the compatibility of the involutions $\imath^\si$ (\ref{010}) and  $\imath^\rho$ (\ref{01}) with non-trivial characteristic classes of underlying bundles.

Assume that the set $\clD$ of  marked points belongs to the invariant set $\Si_g^0$
of the anti-holomorphic involution $\jmath$ of $\Si_{g,n}$.
Consider the one point case $\clD=x$ and the local coordinate $t$ on the disc $D=\{t\,|\,|t|\leq 1\}$
 is $t=z-x$.
%
The involution $\jmath$ acts as $t\to \bar t$.
The representation (\ref{ma}) defines the gluing function
$$
t^{\ga}k(t)=t^{\ga}(k_0+k_1t+\ldots)\,.
$$
The involutions $\imath^\rho$ (\ref{01}) and $\imath^\si$ (\ref{010})
 act on $k(t)$ as
$k(t)\to \jmath^*\rho(k(t)$ and $k(t)\to \jmath^*\si(k(t)$ and on the
multiplier as $t^\ga\to t^{-\ga}$.
It means that the fixed point sets
$\clM_{III}$ and $\clM_{IV}$ of the moduli spaces  $\clM_I$ and $\clM_{II}$ exist
 in the case  $2\ga\in Q^\vee$ only (see Proposition 6 in \cite{BS1}).
It follows from Table 4 in the Appendix B that it happens for the $\SLN$-bundles ($N$ is even), for the bundles
 with the remaining classical gauge groups and for the E$_7$-bundles.

 In the general case, for $\clD=(x_1,\ldots,x_n)$ there are $n$ gluing functions $t^{\ga_a}(k_{0,a}+k_{1,a}t+\ldots)$ defined on the contours $\G_a$. We have $n$ conditions $2\ga_a\in Q^\vee$, $(a=1,\ldots,n)$.
 Type III or IV bundles are nontrivial if the sum
 $\left(\sum_a\ga_a\right)\notin Q^\vee$.


\section{Examples}
\setcounter{equation}{0}


\subsection{Elliptic case}

Consider the elliptic curve $\Si_\tau=\mC/(\mZ+\tau\mZ)$ and let $E$ be the $\SLT$ vector bundle
over $\Si_\tau$. We consider one point case. The marked point is $z=0$. Since $\clZ(\SLT)\sim\mZ_2$ there are two types of bundles corresponding to the trivial
and non-trivial characteristic classes. In the parabolic situation the first case corresponds to the two-particle elliptic
Calogero-Moser system and the second case is the $\SLT$ Euler-Arnold top.
Since the dimensions of both phase spaces  equal two, the systems are integrable.
Here we consider their quasi-antisymmetric extensions and their real  reductions.


\subsubsection{Extension of the Euler-Arnold top}

Consider first non-trivial  bundle of type V.
The sections $s\in\G(E)$ of the rank two bundle have the quasi-periodicities
$s(z+1)=\si_3s(z)$, $s(z+\tau)=\si_1 s(z)$, where
 $\si$-matrices are defined below.
Therefore, the Higgs field takes value in the Lie algebra $\sl2$ and satisfy the following
conditions
\beq{e1}
\Phi(z+1)=\si_3\Phi(z)\si_3\,,~~\Phi(z+\tau)=\si_1 \Phi(z)\si_1,.
\eq
In addition
\beq{e0}
Res\,\Phi(z)|_{z=0}=\bfX\,,
\eq
where $\bfX=g^{-1}\zeta g\in\sl2$ and $\zeta^T=\zeta$ (see also (\ref{mcc}) and (\ref{xr})).
 It means that we attach
     to the point $z=0$ the cotangent bundle $T^*\clX_V$ to the space of complex symmetric matrices $\clX_V=$SO$(2,\mC)\setminus\SLT$.
Instead  of this bundle we consider     the cotangent bundle $T^*(\GLT/$O$(N,\mC))\sim \{(\zeta,g)\}$,
where $g\in\GLT$ and $\zeta=\zeta^T$.

Before passing to $T^*\clX_V$ note that the form $\om$ (\ref{sfcb}) defined on
$T^*\GLT$ is invariant under the multiplication $g\to\la g$, $\la\in\mC^*$.
The quotient $\GLT/\mC^*$ is the group P$\GLT$.
The corresponding moment constraint $\tr\,\zeta=0$, generating this action, is $\tr\,\zeta=0$.
The group P$\GLT$ is covered by $\SLT=\{g\in\GLT\,|\,\det\,g=1\}$ (P$\GLT=\SLT/\mZ_2$).
Fixing the $\mC^*$ action as $\det\,g=1$
we come to the symplectic quotient (\ref{cob}):
\beq{gsq}
T^*\GLT//\mC^*=T^*\SLT=\{\zeta,g)\,|\,\tr\,\zeta=0\,,~\det\,g=1\}
\eq
with the form $\om$ (\ref{sfcb})
 Construct the moduli space of the quasi-antisymmetric bundles
$\clM_V(\Si_\tau\setminus 0,\SLT)$.
Consider first the group GL$(2,\mC)$. We use the basis of the Pauli matrices
 $\si_a$ $(a=0,\dots,3)$.
$$
 \si_0=\left(
\begin{array}{cc}
    1 & 0 \\
    0 & 1 \\
  \end{array}
\right)\,,~
 \si_1= \left(
  \begin{array}{cc}
    0 & 1 \\
    1 & 0 \\
  \end{array}
\right)\,,~
\si_2 = \left(
  \begin{array}{cc}
    0 & -\imath \\
   \imath & 0 \\
  \end{array}
\right)\,,~
 \si_3= \left(
  \begin{array}{cc}
    1 & 0 \\
    0 & -1 \\
  \end{array}
\right)\,.
$$
Now pass to
 the cotangent bundle $T^*\clX_V=T^*({\rm SO}(2,\mC)\setminus\SLT)$ (\ref{gsq}).
In this case $g^*=g^T$. Due to (\ref{cav})
 \beq{paq}
 \clP=g^{-1}\zeta (g^T)^{-1}\,,~~\clQ=g^T g
 \eq
  are complex symmetric matrices.
Their decomposition in the basis of the Pauli matrices assumes the form
\beq{pq}
\begin{array}{c}
  \clP(p_0,p_1,p_3)=p_0\si_0+p_1\si_1+ p_3\si_3\,,~~p_j\in\mC\,, \\
\clQ(q_0,q_1,q_3)=q_0\si_0+q_1\si_1+q_3\si_3\,,~~q_j\in\mC\,.
\end{array}
\eq
The condition $\det\,\clQ\neq 0$ implies that $q_0^2-q_1^2-q_3^2\neq 0$.
It follows from (\ref{caf1}) that  $(p_0,p_1,p_3)$ and $(q_0,q_1,q_3)$ are the Darboux coordinates
on the cotangent bundle $T^*($SO$(2,\mC)\setminus$GL$(2,\mC))$.


The passage to $\SLT$ means that $\det\,g=1$. Then
$$
\det\,\clQ=q_0^2-q_1^2-q_3^2=1\,.
$$

Let $\bfX(p,q)=\clP\clQ=g^{-1}\zeta g$.
\beq{bfp}
\bfX=X_0\si_0+X_1\si_1+\imath X_2\si_2+X_3\si_3\,.
\eq
Then
\beq{bfy}
\begin{array}{l}
  X_0=q_0p_0+q_1p_1+q_3p_3 \\
  X_1=q_0p_1+q_1p_0\,,\\
  X_2=q_3p_1- q_1p_3\,,\\
  X_3=q_0p_3+q_3p_0\,.
\end{array}
\eq

Since $\bfX\in$sl$(2)$,  $\,\tr\,\bfX=\tr\,\zeta=0$.
 Therefore, $X_0 =q_0p_0+q_1p_1+q_3p_3=0$.
Thus, we impose two constraints
\beq{svj}
1.\,c_1=q_0^2-q_1^2-q_3^2-1=0\,,~~2.\,c_2= q_0p_0+q_1p_1+q_3p_3=0\,.
\eq
 These constrains
are the second class constraints (recall that the upstairs brackets are canonical $\{p_i,q_j\}=\delta_{ij}$ ):
$$
\{c_1,c_2\}|_{\varphi_1=0}=-2\,,
$$
i.e. the matrix of the Poisson brackets between the constraints $\{c_i,c_j\}$ takes the form
$$
\mat{0}{-2}{2}{0}
$$
on-shell (\ref{svj}).
Taking into account the brackets
$$
\begin{array}{ccc}
  \{p_0,c_1\}=2q_0\,, &\{p_{1,2},c_1\}=-2q_{1,2}\,,  &\{q_j,c_1\}=0\,,\\
  \{p_j,c_2\}=p_j \,,& \{q_j,c_2\}=-q_j  &
\end{array}
$$
we come to the following reduced Dirac brackets on-shell the constraints (\ref{svj}):
  \beq{DB}
  \begin{array}{c}
 \displaystyle{
  \{p_0,q_0\}=1-q_0^2\,,\quad \{p_0,q_1\}=-q_0q_1\,,\quad  \{p_0,q_3\}=-q_0q_3\,,
  }
  \\ \ \\
  \displaystyle{
  \{p_1,q_0\}=q_0q_1\,,\quad \{p_1,q_1\}=1+q_1^2\,,\quad \{p_1,q_3\}=q_1q_3\,,
  }
  \\ \ \\
   \displaystyle{
  \{p_3,q_0\}=q_0q_3\,,\quad \{p_3,q_1\}=q_1q_3\,,\quad \{p_3,q_3\}=1+q_3^2\,.
  }
 \end{array}
 \eq
 These brackets along with the constraints (\ref{svj}) defines the Poisson structure
 on the cotangent bundle $T^*\clX_V=T^*({\rm SO}(2,\mC)\setminus\SLT)$.

It  can be checked directly from (\ref{bfy}) that
the coefficients $X_\al$ satisfies the Poisson-Lie sl$(2)$ algebra
\beq{pla}
\{X_\al,X_\be\}=c_{\al\be\ga}X_\ga\,,~~(\al\,,\be\,,\ga=1,2,3)\,,
\eq
or in details
$$
\{X_1,X_2\}=-X_3\,,~~\{X_2,X_3\}=- X_1\,,~~\{X_3,X_1\}= X_2\,.
$$
The Casimir function of this algebra is
\beq{cas}
\clC=\tr\,\bfX^2=X_1^2-X_2^2+X_3^2\,.
\eq
It should be emphasized that $\clC$ is not the Casimir function of the Poisson algebra
on \\
$T^*($GL$(2,\mC/$SO$(2,\mC))$, generated by
 $(p_j,q_k)$ ($j,k=0,1,3)$.


For the $\SLT$ Higgs bundles over the elliptic curve $\Si_\tau$ with the quasi-periodicities
(\ref{e1}) one can take $\bA=0$ as the gauge fixing condition. After the gauge transformation
the Higgs field becomes the Lax operator $L_V$.
 It satisfies the holomorphicity condition
 (\ref{mome}) and has a first order pole at $z=0$.
 The moduli space is defined by the residue $res\,L(z)|_{z=0}=\bfX$
 \beq{m5}
 \clM_V(\Si_\tau\setminus \{0\},\SLT)\sim (T^*\clX_V=T^*(\SLT/{\rm SO}(2,\mC))=
 \eq
 $$
 \{((p_0,q_0),(p_1,q_1)(p_3,q_3))\,|\,(c_1=0\,,\,c_2=0)\}\,.
 $$
 It has complex dimension 4.

The Lax operator can be defined in terms of the Kronecker function.
The Kronecker function $\phi(u,z)$ is related to the elliptic curve $\Si_\tau$
and takes the form
\beq{phi}
 \phi(u,z)=\frac{\vth(u+z)\vth'(0)}{\vth(u)\vth(z)}\,,
 \eq
 where $\vth(z)$ is  the theta-function
\beq{theta}
\vth(z|\tau)=q^{\frac
{1}{8}}\sum_{n\in {\bf Z}}(-1)^ne^{\pi i(n(n+1)\tau+2nz)}\,.
\eq
The Kronecker function has the following  quasi-periodicities:
\beq{A.14}
\phi(u,z+1)=\phi(u,z)\,,~~~\phi(u,z+\tau)=e^{-2\pi \imath u}\phi(u,z)\,,
\eq
and has the first order pole at $z=0$
\beq{rk}
\phi(u,z)\sim\f1{z}\ldots\,.
\eq
It is related to the Weierstrass function $\wp$ as follows:
\beq{wpphi}
\phi(u,z)\phi(-u,z)=\wp(z)-\wp(u)\,.
\eq
Let
$$
\varphi_1(z)=\phi(\oh,z)\,,~~\varphi_2(z)=\exp(\pi\imath z)\phi(\frac{1+\tau}2,z)\,,~~
\varphi_3(z)=\exp(\pi\imath z)\phi(\frac{\tau}2,z)\,.
$$
Then from  (\ref{e1}), (\ref{e0}),   (\ref{A.14}) and (\ref{rk}) we find the Lax operator
\beq{cl}
L_V=X_1\varphi_1(z)\si_1+X_2\varphi_2(z))\si_2+
X_3\varphi_3(z))\si_3\,,
\eq
where $X_\al$ are given by (\ref{bfy}).
Since $\tr\, L^2$ is a double periodic function it can be expressed in terms of  the
$\wp$-function. By means of (\ref{wpphi}) define the expansion
$\tr\, L_V^2=2H^V_{2}\wp(z)+2H^V_0$,
where the coefficients are two Hamiltonians
\beq{Ha2}
H^V_2=\oh(X^2_1-X_2^2+X^2_3)=
\eq
$$
=\oh\Big((q_0p_1+q_1p_0)^2-(q_1p_3-q_3p_1)^2+
(q_0p_3+q_3p_0)^2\Big)\,,
$$
\beq{Ha0}
H^V_0=\oh(X^2_1\wp(\oh)-X^2_2\wp(\frac{1+\tau}2)+ X^2_3\wp(\frac{\tau}{2})=
\eq
$$
=\oh(q_0p_1+q_1p_0)^2\wp(\oh)-\oh(-q_1p_3+q_3p_1)^2\wp(\frac{\tau+1}{2})+
\oh(q_0p_3+q_3p_0)^2\wp(\frac{\tau}{2})\,.
$$
Since $H^V_2$ (\ref{Ha2}) is proportional to the Casimir  function (\ref{cas}), the Hamiltonians
$H^V_2$ and $H^V_0$  Poisson commute.

We replace the Hamiltonian $H^V_0$ by the Hamiltonian $\ti H^V_0$, where the values of the $\wp$-functions
are replaced by an arbitrary complex coefficients
\beq{h5}
\ti H^V_0(J_1,J_2,J_3)=\oh( X^2_1 J_1- X^2_2 J_2+ X^2_3 J_3)=
\eq
$$
=\oh(q_0p_1+q_1p_0)^2 J_1-\oh(-q_1p_3+q_3p_1)^2J_2+
\oh(q_0p_3+q_3p_0)^2J_3\,.
$$
It commutes with $H^V_2=\ti H^V_0(1,1,1)$ as before.

These Hamiltonians being restricted on the phase space
$T^*(\SLT/$SO$(2,\mC))$ also Poisson commute, because
the symplectic reduction (\ref{gsq})
does not break the involutivity of the Hamiltonians.
For  $J_k\neq 1$ they are independent. Therefore, we come to the complex completely
integrable system on $T^*(\SLT/$SO$(2,\mC))$, because $\dim\,(\SLT/$SO$(2,\mC))=2$.
The system is described by the two commuting  Hamiltonians $\ti H^V_0$, $H^V_2$, the brackets (\ref{DB}) and constraints (\ref{svj}).

Consider the Hamiltonian $\ti H^V_0$ and the corresponding equations of motion
 before the passage to the group $\SLT$.
Using (\ref{DB}) we obtain
\beq{eq5}
\begin{array}{ll}
  \dot q_0=J_1X_1q_1+J_3X_3q_3\,, & \dot p_0=-J_1X_1p_1-J_3X_3p_3\,, \\
\dot q_1=J_1X_1q_0- J_2X_2q_3  \,,& \dot p_1=-J_1X_1p_0-J_2X_2p_3  \,,\\
 \dot q_3= J_2 X_2q_1+J_3X_3q_0 \,, & \dot p_3=-J_3X_3p_0+ J_2 X_2p_1 \,.
\end{array}
\eq
On the other hand since the variables $X_\al$ (\ref{bfy}) satisfy the sl$(2)$ Poisson algebra (\ref{pla}),
$\ti H^V_0$ (\ref{h5}) is the Hamiltonian of the Euler-Arnold top on the GL$(2,\mC)$
orbit corresponding to the value of $H^V_2$ (\ref{Ha2}). The equations of motion are of the form:
$$
\dot X_\al =-c_{\al\be\ga}J_\be X_\be X_\ga \,.
$$
Comparing these equations with (\ref{eq5}) we conclude that
 $X_\al(p_0,\ldots,q_3)$ play the role of collective coordinates on the phase space
$T^*({\rm SO}(2)\setminus$GL$(2,\mC))$. As in the general case the coadjoint orbit arises here as
the symplectic quotient of the cotangent bundle $T^*({\rm SO}(2)\setminus\SLT)$ with respect to
the action of the diagonal subgroup of $\SLT$.


 \subsubsection{Real reductions}

 Consider the real symmetric spaces
 $$
 \clX_{III}=S^2={\rm SO}(2)\setminus{\rm SU}(2)\,,~~
\clX_{IV}={\rm SO}(2)\setminus{\rm SL}(2,\mR)\,.
 $$
 Here $\clX_{IV}$ is the Lobachevsky plane.

In the first case in (\ref{bfy}) $(q_0,p_0)\in\mR$, $(q_1,p_1),(q_3,p_3)\in\imath\mR$. The coefficients $X_\al\in\imath\mR$ and
satisfy the su$(2)$ Poisson-Lie algebra
$$
\{X_\al,X_\be\}=-\imath\ep_{\al\be\ga}X_\ga\,.
$$
The sum
$$
H^{III}_2=-\oh(X^2_1+X_2^2+X^2_3)
$$
is positive. It is the Casimir function of this algebra.

In the second case all coordinates and momenta $(p_j,q_j)$ are real. Then the coefficients $X_1$, $X_3$ are real and $X_2$ is imaginary. The Poisson brackets coincide
with (\ref{pla}). Consequently, the Casimir function, which will play the role of the Hamiltonian
assumes the form
$$
H^{IV}_2=\oh(X^2_1-X_2^2+X^2_3)\,.
$$
This Casimir function is real.

Our goal is to construct two Lax operators as a result of involutions (\ref{io1}) and
(\ref{io2}).
To define the involution of the $\SLT$ Higgs bundles we need
 elliptic curve with anti-holomorphic involutions $\jmath$.
 If   $\tau=\imath t$, $t\in\mR^+$ then the involution is conjugation $\jmath\,(z)=\bz$.
The fixed point set $\Si^0_{\imath t}$  is $\Re e\,z=x$, or $z=\oh \imath t+x$, where $0\leq x<1$.
Another choice of the fixed point set is $z=a+\imath x$, $a=0$, or $a=\oh$.
The coefficients $\wp(\oh)$, $\wp(\frac{1+\imath t}2)$, $\wp(\imath t/2)$
of the Hamiltonian $H^V_0$ (\ref{Ha0}) belong to the fixed points set. They are real.

Another possibility is to define the elliptic curve as the quotient $\Si_{\om_1\om_2}=\mC/(2\om_1\mZ+2\om_2\mZ)$, $(\Re e\,\om_j>0)$
and $\om_2=\bar\om_1$.
We will not analyze here this case.
Real involutions for many marked points on elliptic curves and integrability
were considered in \cite{GD}.

The real Higgs SU$(2)$-bundles (for the system of type III) and
 SL$(2,\mR)$-bundles (for the system of type IV) are results of reductions
(\ref{dia}) of $\clM_V$ (\ref{m5}). We have two types of the moduli spaces
\beq{m3}
\clM_{III}(\Si^0_{\imath t}\setminus 0,{\rm SU}(2))\sim
(T^*\clX_{III}=T^*({\rm SO}(2)\setminus{\rm SU}(2))=
 \eq
 $$
 \{((p_0,q_0)\in\mR,(p_1,q_1)(p_3,q_3)\in\imath \mR)\,|\,(c_1=0\,,\,c_2=0)\}\,.
 $$
\beq{m4}
\clM_{IV}(\Si^0_{\imath t}\setminus 0,{\rm SL}(2,\mR))\sim
(T^*\clX_{IV}=T^*({\rm SO}(2)\setminus{\rm SL}(2,\mR))=
 \eq
 $$
 \{((p_0,q_0)(p_1,q_1)(p_3,q_3)\in \mR)\,|\,(c_1=0\,,\,c_2=0)\}\,.
 $$
 The Lax operators (the Higgs fields) assume the forms
\beq{cl1}
L^{III,IV}=X_1\varphi_1(x)\si_1+X_2\varphi_2(x))\si_2+
X_3\varphi_3(x))\si_3\,,~~x\in\Si^0_{\imath t}\,,
\eq
where $X_\al$ are imaginary for the systems of type III and $X_1,X_3$ are real and $X_2$ are
imaginary for the systems of type IV.

As before we come to the Hamiltonians
$$
\ti H^{III}_0=\oh\sum_\al X_\al^2J_\al\,,~~\ti H^{IV}_0=\oh( X_1^2J_1-X_2^2J_2+X_3^2J_3)
$$
commuting with $H^{III}_2$ and $H^{IV}_2$.

 The system of type III corresponds to a motion of a particle on the two-dimensional
 sphere $S^2$ and of type IV to a motion on the Lobachevsky plane.
 The moduli spaces
  $\clM_{III}(\Si_{\imath t}^0\setminus 0,{\rm SU}(2))=T^*S^2$ (see \ref{m5})) is the
 Eguchi-Hanson gravitational instanton \cite{EH}. It has the hyper-Kahler structure \cite{Hi2}.
 In the general case we did not find the hyper-Kahler structure on the space $\clM_{V}(\Si_\tau\setminus \{0\},\SLT)$.

 Consider the corresponding quantum system.
 The Hamiltonian $\ti H_0^V$ (\ref{h5}) is quantized by the canonical quantization rules
 and subsequent imposing of (the quantization) constraints (\ref{svj}). The quantization in terms of
 the generators $\hat X_\al$ of the algebra su$(2)$ is straightforward \cite{Vi}.
 The generators act on the space of functions on $S^2$ that are quadratically integrable with respect to the invariant measure.
 The Hamiltonian $\hat H^{III}_2$
 is the Laplace-Beltrami operator defined by the su$(2)$ invariant metric on $S^2$.
 In the spherical coordinates  on $S^2$ the operator $\hat H^{III}_0$  assume the form
 \beq{sco}
 \hat H^{III}_0=(\cos\psi\p_\te-\tan^{-1}\te\sin\p_\psi)^2J_1+
 (-\sin\psi\p_\te-\tan^{-1}\te\cos\psi\p_\psi)^2J_2 +\p^2_\psi J_3\,.
 \eq
  The eigen-subspaces $V_l$ of $H^{III}_2$ are irreducible
 unitary representations of su$(2)$ of spin $l$, ($l=0,\oh,1,\ldots$) and $\dim\,V_l=2l+1$.
 The basis of $V_l$ is generated by the functions
 $$\Psi_{lm}=P_l^m(\cos\te)\exp\,(-\imath m\psi)\,,\quad -l\leq m\leq l,$$
 where $P_l^m(\cos\te)$ are the associated Legendre polynomials.
 Since $\hat H^{III}_2$ and $\hat H^{III}_0$ commute we can restrict the latter operator
 on the space $V_l$.
 Therefore, the  eigenfunctions of $\hat H^{III}_0$ are the linear combination of the functions $\Psi_{lm}$.



\subsection{Extension of Calogero-Moser system}

Consider the trivial type V SL$(2,\mC)$ Higgs  bundle over $\Si_\tau$.
 The section of the bundle have the quasi-periodicities
(compare with (\ref{e1}))
$$
\Phi(z+1)=\Phi(z)\,,~~\Phi(z+\tau)=Q(u)\Phi(z)Q^{-1}(u)\,,~~
Q(u)=\di(\bfe(u),\bfe(-u))\,,~(\bfe(u)=\exp\,2\pi\imath u)\,.
$$
By the gauge transforms the connection can be trivialised $\bA=0$.
The system has residual symmetry acting on the variables as in (\ref{fgt}), (\ref{ht}) by constant
diagonal matrices. This action produces the moment constraint $\bp\Phi|_{\di}=0$.
Therefore the diagonal part of the residue $\bfX$ (\ref{e0}) vanishes. In this way we come to the
constraint  (see (\ref{bfy}))
\beq{ac}
X_3=0\,,~~ (q_0p_3+q_3p_0=0)\,.
 \eq

Let $\si_{\pm}=\oh(\si_1\pm\imath\si_3)$ and $X_\pm=(X_1\pm X_2)$.
 The Lax operator assumes the form
 \beq{Lac}
 L^V(z)=v\si_3+X_{\pm}\bfe(\pm 2u)\phi(\pm 2u,z)\si_{\pm}\,.
 \eq

  Expanding as above $\tr\,(L^2(z))$ we find two Hamiltonians
  \beq{hc2}
 2 H^V_2=X_+X_-=X_1^2-X_2^2\,,
  \eq
  \beq{hc0}
  H^V_0=\oh v^2+X_+X_-\wp(2u)\,,
  \eq
  where $(v,u)$ are the canonical variables in the Calogero-Moser system.
  Fix the value of $H_2=\mu^2$.
  Due to (\ref{ac}) the Hamiltonian $H_2$ is proportional to the Casimir function (\ref{cas}).
 It plays the role of the coupling constant
in the Hamiltonian $H_0$. The constant has the internal degrees
of freedom
$$
X_+X_-=(q_0p_1+q_1p_0)^2-(-q_1p_3+q_3p_1)^2\,.
$$
Thereby, the variables $(p_j,q_j)$ have a nontrivial dynamic with respect to
the Hamiltonian $H_0$. This dynamic depends on the Calogero-Moser  coordinate $u$,
while the dynamic of the Calogero-Moser system depends on the value of
Hamiltonian $H_2$ only.

The reductions to the type III and IV bundles lead to the corresponding generalizations
of the Sutherland type systems
$$
 H^{III}_0=\oh v^2+\frac{X_+X_-}{\sinh^2(2u)}\,,~~\,\, H^{IV}_0=\oh v^2+\frac{X_+X_-}{\sin^2(2u)}\,.
$$


\subsection{Rational case}

Consider $\SLT$ quasi-antisymmetric Higgs bundles over $\mC P^1$ with $n$ marked points $\clD=(x_1,\ldots,x_n)$.
The moment equation (\ref{mome}) takes the form
\beq{rm}
\bp L(z,\bz)=\sum_{a=1}^n\de(z-x_a,\bz-\bar x_a)\bfX_a\,,
\eq
where (see (\ref{bfp}))
$$
\bfX_a=X^a_1\si_1+X^a_2\si_2+X^a_3\si_3\,.
$$
The moduli space is a result of the symplectic reduction
$$
\clM_V(\mC P^1\setminus\clD,\SLT)=\SLT\setminus\setminus(\prod_aT^*\clX{V,a})\,.
$$
Thereby the dimension of the moduli space is
 $\dim\,\clM_V(\mC P^1\setminus\clD,\SLT)=2(2n-3)$.

As above we consider first the GL$(2,\mC)$ bundles.
 Let $\bfX_a=\clQ^a\clP^a$
where (see (\ref{pq}))
\beq{pq1}
\begin{array}{c}
  \clP^a(p^a_0,p^a_1,p^a_3)=p^a_0\si_0+p^a_1\si_1+ p^a_3\si_3\,,~~p^a_j\in\mC\,, \\
\clQ^a(q^a_0,q^a_1,q^a_3)=q^a_0\si_0+q^a_1\si_1+q^a_3\si_3\,,~~q^a_j\in\mC\,.
\end{array}
\eq
Then
\beq{Xa}
\begin{array}{l}
    X^a_1=q^a_0p^a_1+q^a_1p^a_0\,, \\
    X^a_2=-q^a_1p^a_3+q^a_3p^a_1\,, \\
    X^a_3=q^a_0p^a_3+q^a_3p^a_0\,.
  \end{array}
\eq
As above $(p^a_j,q^b_k)$ ($j,k=0,1,3)$ are the Darboux coordinates
$\{p^a_j,q^b_k\}=\de^{ab}\de_{jk}$.
As a consequence for the collective coordinates we have (see (\ref{pla}))
\beq{gb}
\{X^a_\al,X^b_\be\}=\de^{ab}c_{\al\be\ga}X_\ga\,.
\eq
The Lax operator
$$
 L(z,\bz)=\sum_{a=1}^n\frac{\bfX_a}{z-x_a}
$$
 is a solution of the equation (\ref{rm}).
Since
$$
\tr\,( L^2)=\sum \frac{(\bfX^2_a)}{(z-x_a)^2}+\left(\sum_{b\neq a}\frac{(\bfX_a\bfX_b)}{x_b-x_a}
\right)\f1{z-x_a}
$$
we come to Gaudin type Hamiltonians:
$$
H_2^a=(\bfX^2_a)\,,~~
H_1^a=\sum_{b\neq a}\frac{(\bfX_a\bfX_b)}{x_b-x_a}\,.
$$
The Hamiltonians $H_2^a$ are the Casimir functions with respect to the brackets (\ref{gb}),
and thereby they commute with the Hamiltonians $H_1^b$.
But they are not the Casimir functions in the Poisson algebra defined on the phase space
$\prod_a\,T^*($GL$(2,\mC/$SO$_a(2,\mC))$.
  In terms of $\bfX_a$ ($a=1,\dots,n$) $L(z)$ is the standard Lax operator for the Gaudin system.
The latter is contained in the phase space $\clM_V(\mC P^1\setminus\clD,\SLT)$.
The number of independent integrals is $2n-3$, while for the Gaudin system it is equal to $n-3$.

Let $\p_{a,2}=\{H_2^a,~\}$ and $\p_{a,1}=\{H_1^a,~\}$.
In terms of the Darboux matrices $\clP_a$, $\clQ^a$ (\ref{pq1}) the equations of motion take the
form
$$
\begin{array}{l}
  \p_{a,2}\clQ^b=\de^{ab}(\clQ^a\clP^a\clQ^a+(\clQ^a)^2\clP^a)\,, \\
  \p_{a,2}\clP^b =-\de^{ab}(\clQ^a(\clP^a)^2+\clP^a\clQ^a\clP^a)\,,\\
  \p_{a,1}\clQ^b =\f1{x_b-x_a}\clQ^a\clP^a\clQ^b\,, \\
  \p_{a,1}\clQ^a=\sum_{b\neq a}\f1{x_b-x_a}\clQ^a\clQ^b\clP^b\,, \\
  \p_{a,1}\clP^b =\f1{x_a-x_b}\clQ^a\clP^a\clP^b\,, \\
  \p_{a,1}\clP^a = \sum_{b\neq a}\f1{x_a-x_b}\clP^a\clQ^b\clP^b\,
\end{array}
$$
Consider the anti-involution $\jmath(z)=\bz$ of $\mC P^1$. The marked points of the invariant set $$(x_1,\ldots,x_n)\in\clD$$ are
either real $x_a\in\mR$, or combined in the conjugated pairs $x_a=\bar x_{a+1}$.

For the systems of type III the residues satisfy conditions
$$
\begin{array}{l}
  \bfX_a=-\bar\bfX_a~{\rm for~}x_a\in\mR\,, \\
   \bfX_a=-\bar\bfX_{a+1}~{\rm for~}x_a=\bar x_{a+1}\,.
\end{array}
$$
The case when all $x_a\in\mR$ corresponds to the real reduction of the $\SLT$ Gaudin
system.

For the systems of type IV the residues satisfy conditions
$$
\begin{array}{l}
  \bfX_a=\bar\bfX_a~{\rm for~}x_a\in\mR\,, \\
   \bfX_a=\bar\bfX_{a+1}~{\rm for~}x_a=\bar x_{a+1}\,.
\end{array}
$$
In both cases the moduli spaces has the same structures
as in (\ref{m3}) and (\ref{m4}), where the coordinates and momenta acquire
the additional index $"a"$ related to the position of the marked point.


\section{Conclusion}

\setcounter{equation}{0}

The structures described in this paper are related to the $\clN=4$  SUSY Yang-Mills theory
on a four-dimensional space \cite{KW,GW,GW2}.

First, the Higgs bundles are related to the Hitchin equations that can be defined in the following
way. Let $(\Phi,\bA)$ be the Higgs data and $\rho$ is the involutive automorphism
$\rho$ of $G^\mC$ with the fixed point set to be the maximal compact subgroup $C$ of $G^\mC$
(\ref{rho}).
Using the connection $d_{\bA}$  define the unitary
connection $\nabla_A=d_{\bA}+d\rho(d_{\bA})$, $(d\rho(d_{\bA})=\p_z+d\rho(\bA))$
\beq{A}
d_{\bA}=(\p_{\bz}+A_{\bz})d\bz\,,~d_A=(\p_z+A_z)dz\,,~~A_{\bz}=\bA\,,~A_z=d\rho(\bA)\,,
\eq
where $d\rho$ is the corresponding  involutive automorphism acting on the Lie algebra $\gg^\mC$.

Similarly, for the Higgs field $\Phi$ define the
adjoint field $\Phi^*=-d\rho(\Phi)$ and
\beq{uh}
\phi=\phi_zdz+\phi_{\bz}d\bz\,,~~\phi_z=\Phi\,,~\phi_{\bz}=\Phi^*\,.
\eq
In these terms the Hitchin equations assume the form \cite{Hi}
\begin{subequations}\label{heq}
  \begin{align}
& F(A)-\phi\wedge\phi=0 \,,\\
 &d_A\phi_{\bz}+d_{\bA}\phi_z=0\,,\\
  &\imath(d_A\phi_{\bz}-d_{\bA}\phi_z)=0 \,.
\end{align}
\end{subequations}

The gauge group
$\clG(\Si_g,C)= \,Map^{C^\infty}(\Si_g\to C)$
acts on the space of solutions of the Hitchin equations. Let
$$
\gM(\Si_g,C)=
\clG^C\setminus\!\{\rm solutions~of~(\ref{heq})\}\,.
$$
be the moduli space. This construction endows $\gM(\Si_g,C)$ with the \ hyper-Kahler
structure \cite{HKR}. Thereby, $\gM(\Si_g,C)$ has a $CP^1$ family of complex structures.

The last two equations in (\ref{heq}) can be combined into the complex form
$d_{A_{\bz}}\phi_z=0$. This equation coincides
with the moment equation for the Higgs bundles (\ref{mome}) with $n=0$. Recall that
the moduli space of the Higgs bundles is the quotient with respect to the action of
the complex gauge group
$$
\clM(\Si_g,G^\mC)=\{d_{A_{\bz}}\phi_z=0\}/\clG(\Si_g,C)\,.
$$
In one of these complex structures $I$ the moduli space $\gM(\Si_g,C)$ is isomorphic to a smooth
 part of the Higgs moduli space $\clM^0(\Si_g,G^\mC)$ \cite{Hi,Si1}.

On the other hand, the Hitchin equations (\ref{heq})  arise in the
topological version of the twisted
$\clN=4$ SUSY  Yang-Mills theory on the four-dimensional space-time $M$ \cite{KW}.
Let $M=\mR^2\times\Si_{g,n}$.
There is a reduction of this four-dimensional gauge theory to the two-dimensional
sigma-model defined by the map of $\mR^2$ to the moduli space $\gM(\Si_{g},C)$.

The moduli space $\gM(\Si_{g,n},C)$ of the Hitchin equations defined on multipoint curves
 is a result of including in the four-dimensional gauge theory two-dimensional
 surface operators \cite{GW}. Let $\ti D$ be a two-dimensional surface $\ti D\subset M$
  that  intersects transversally with the curve $\Si_{g,n}$ along
 the set of the marked points $\ti D\cap\Si_{g,n}=\clD(x_1,\ldots,x_n)$.
 As it is explained in \cite{GW2} the parabolic Higgs bundles over multipoint curves
are described by the interactions of the $\clN=4$ Yang-Mills theory with a sigma-model supported on
the surface $\ti D$. The sigma-model on  $\ti D$ is called the surface operator.
For the parabolic bundles the target spaces of the sigma-model
are the co-adjoint $G^\mC$-orbits. These spaces are hyper-Kahler and
the total space $\gM_{par}(\Si_{g,n},C)$ is also hyper-Kahler \cite{GW}.
 It was established in \cite{Kr1}
 that the cotangent bundles $T^*G^\mC$  is a hyper-Kahler
 variety. Apparently, the whole theory describing the interaction
 of the $\clN=4$ Yang-Mills theory with a sigma-model with the target spaces $T^*G^\mC$
 is not hyper-Kahler, but only Kahler.
 This phenomenon is most likely to occur in the case of
  $T^*\clX_V$. It means that including the surface operators of these types
 breaks the $\clN=4$ SUSY  to $\clN=2$.

%

It was found in \cite{BS}
 that
the fixed point sets of the real involutions (\ref{io}) are  special branes $\clB$
in the space $\clM^0(\Si_{g},G^\mC)$ corresponding to different complex structures in
 $\gM(\Si_{g},G^\mC)$.
We can apply the same involutions to the parabolic bundles and come to integrable
systems. Let $H_a$ be the Cartan subgroup of $G^\mR$ and $T_a$ is the Cartan torus of
the compact group $C$. The spin variables belong to the following
 types of orbits. The real orbits  $H_a^\mR\setminus G^\mR$
 are attached to  the marked points of the fixed point set of the involution $\iota^\si$ and
the flag varieties $T_a\setminus C$
are attached to  the marked of the fixed point set of the involution $\iota^\rho$.

\begin{center}
\textbf{APPENDIX}
\end{center}
\appendix


\section{Two equivalent constructions of symplectic quotients}

 Let $H$ be a subgroup if a group $G$ and $X$ be a $G$-set.
We consider two quotient spaces:
the \textbf{$\al$-model} $(X\times H\backslash G)\slash G$ and the
\textbf{$\be$-model} $ X/ H$
 Our aim is to prove they are isomorphic.
\begin{predl}
The $\al$ and the $\be$ models are isomorphic
 ($ X/H\sim (X\times H\backslash G)/G$).
 \end{predl}
 \emph{Proof}\\  First we construct a map
$$
\zeta:  X/H\to  (X\times  H\backslash G)/G\,.
$$
 Denote by $[g]$ the coset $Hg$.
The map
$$
\tilde{\eta}\,:\,X\to X\times H\backslash G\,,~x\to (x,[1_G])
$$
 is $H$-invariant.
Hence we have the induced map
$$
\eta:  X\slash H=  (X\times H\backslash G)\slash H\,.
$$
The required map $\al$ is  composition of $\eta$ with the natural projection $\zeta= \pi\circ\eta$
$$
\pi\,:\,(X\times G\slash H)\slash  H\to  (X\times H\backslash G)\slash G\,.
$$

Second we prove that $\zeta$ is injective. Let $\zeta(x)=\zeta(y)$.
So $(y,[1_G])$ belongs to $G$ orbit of $(x,[1_G])$:
$(y,[1_G])=(g(x)x,g([1_G]))$ for some $g\in G$. But $g([1_G])=[g]$
 so $[g]=[1_G]$ and $g\in H$. We get $y=g(x)\in Hx$.

Third we prove surjectivity. Any element $(X, [g])$ in $X\times G\slash H$
 belongs to $G$ orbit of the element
$g^{-1}(X, [g])=(g^{-1}(X), [1_G])\in \eta(X)$.
$\Box$

Let $ G$ and $H$ be Lie groups and $M$ is a symplectic manifold with Hamiltonian action of $G$.
\begin{predl}
$M\slash\slash_\nu H  =  (M\times H\backslash\backslash_{\nu} T^*G)\slash\slash G$
 \end{predl}

 If $\mu\in$Lie$^*(H)$ is the moment map defining the symplectic reductions then
in the notations $\backslash\backslash_\nu$, $\slash\slash_{\nu}$ $\nu$ means the value
of $\mu$.

 \emph{Proof.}
  We start from construction of a map.
Denote by $\mu_G$ ($\mu_H$) the momentum map to Lie coalgebra ${\frak g}^*$ (${\frak h}^*$).
 Since ${\frak h}\subset{\frak g}$
one has  dual projection  $\pi: {\frak g}^*\to{\frak h}^*$.
As the action of $H$ is the restriction of the action of $G$,
$\mu_H=\pi \circ \mu_G$

The total space of the cotangent bundle $T^*G$ equals to $G\times {\frak g}^*$
and the right action of the group on itself induces coadjoint action ${\rm Ad}^*$ on the  coalgebra.
The momentum of this action is just the opposite to identical map on the second factor.
Hence  an element
of  $H\backslash\backslash_{\nu}T^*G$ can be described as a pair $([g],\theta)$, $[g]$ be a coset as above and $\theta\in {\frak g}^*$ such that
$\pi (\theta)=-\nu$.

The Hamiltonian quotient $  M\slash\slash_\nu H$
is the quotient by action of $H$ of the momentum level $M^\nu=\mu_H^{-1}(\nu)$.

Define the map
$$
\tilde{\eta}_H: M\to M\times H\backslash\backslash_{\nu}T^*G \,:
 X\to (X,([1_G],-\mu_G(X)))\,.
 $$
  Since $\pi(-\mu_G(X))=- \mu_H(X)=-\nu$ the second component
  $([1_G],-\mu_G(X))$ lies in $H\backslash\backslash_{\nu} T^*G H$.

The $G$-momentum of an element $(X,([g],\theta))$ in the product
$M\times H\backslash\backslash_{\nu} T^*G $
 is equal to
$\mu_G (X)+\theta$. So the image of the map $\tilde{\eta}_H$ lies in the zero
$G$-momentum level $ (M\times H\backslash\backslash_{\nu} T^*G )^{\mu_G=0}$.

Evidently, the map $\tilde{\eta}_H$ is $H$-invariant.
Hence we have induced map
$$
\eta_H: H M^\nu\slash H = (M\times H\backslash\backslash_{\nu} T^*G )^{\mu_G=0}\slash H\,.
$$
  Its composition with natural projection
  $$
H\slash (M\times H\backslash\backslash_{\nu}T^*G)^{\mu_G=0}
\to  G\slash (M\times H\backslash\backslash_{\nu} T^*G )^{\mu_G=0}
$$
 is the required map $\zeta_H$.

Injectivity of the map $\zeta_H$ follows
from the fact that the stabilizer of the trivial coset $[1_G]$ is exactly $H$,
so the proof is very close to the proof of injectivity in Prop 1.

For surjectivity we (again like in Prop 1) we choose in $G$- orbit a representative
$(X,([1_G],\theta)) $ with trivial coset as component.
As the $G$-momentum $\mu_G(X)+\theta$ should vanishes, $\theta= -\mu_G(X)$,
hence this element is $\tilde{\eta}_H(X)$.
$\Box$
\begin{predl}
 The isomorphism $\zeta_H$ from the previous proposition is symplectic.
 \end{predl}
 \emph{Proof}\\
 The symplectic form on the symplectic quotient is uniquely determined by the following property:
its pull-back to the momentum level coincides with restriction of the initial symplectic form.

Since the restriction of the symplectic form on $T^*G$ to the fiber ${\frak g}^*$ vanishes,
the map $\tilde{\eta}_H$ is symplectic. Since $\zeta$  is injective, a
$G$-orbit in  $M\times H\backslash\backslash_{\nu}T^*G $ cut on the image of $M$ the image under
 $\tilde{\eta}_H$  of some $H$-orbit in $M$ on the image of $M$.
 Hence the pullback of the restriction   of the form to
 $ (M\times H\backslash\backslash_{\nu}T^*G )^{\mu_G=0}$ coincides with
the restriction to $M^\nu$ and the corresponding projections to Hamiltonian quotients are equal.
These forms are pullback under projections of the symplectic  forms on the Hamiltonian quotients,
hence this symplectic forms coincides.
$\Box$


\section{Simple Lie groups: notations and decompositions }
\setcounter{equation}{0}

Let $G^\mC$ be a simple  complex Lie group, $\gg^\mC$ its Lie
algebra of rank $l$ and $\gh^\mC$ is a Cartan subalgebra.

\paragraph{Chevalley basis in $\gg$.}
Let $\{\al\in R\}$ be the root system, and $R^\pm$ are subsystems of positive
(negative) roots with respect to some ordering.

The algebra $\gg^\mC$ has the root basis
 \beq{CB}
\{E_\al\,,\,\al \in R\}\,,\,\{e_1,\ldots,e_l\},,
 \eq
 where $(e_j)$ is a basis in $\gh^\mC$.

There are two sets of generators  we use here:
\beq{cb}
\{(E_\al-E_{-\al})\,,\, \imath(E_\al+E_{-\al})\,,~\al \in R^+\}\,,\,\{\imath e_1,\ldots,\imath e_l\}\,.
 \eq
\beq{ab}
\{(E_\al-E_{-\al})\,,\, (E_\al+E_{-\al})\,,~\al \in R^+\}\,,\,\{ e_1,\ldots, e_l\}\,.
 \eq

 \noindent
\emph{\textbf{Centers of simple groups}}.\\
 Consider the dual root system $R^\vee=\{\al^\vee=\frac{2\al}{(\al,\al)}\,|\,\al\in R$,
and let $Q^\vee$ be the coroot lattice in $\gh^\mC$. The coweight lattice $P^\vee\subset\gh^\mC$
is the dual lattice to the root lattice $Q$ generating by $R$.
Let  $G^\mC$ be an universal covering group. The group $G^\mC$ is simply-connected
and in all cases apart $G_2$, $F_4$ and $E_8$ has a non-trivial center $\clZ(G^\mC)$.
The center is isomorphic to the quotient group $P^\vee/Q^\vee$.

\begin{center}

\begin{tabular}{|c|c| }
  \hline
\hline
   $G^\mC$  & $\clZ(\bar{G})$ \\
 \hline
 \hline
SL$(n,\mC)$  & $\mu_n$  \\
Spin$_{2n+1}(\mC)$ & $\mu_2$  \\
Sp$_n(\mC)$ & $\mu_2$   \\
Spin$_{4n}(\mC) $& $\mu_2\oplus\mu_2$     \\
Spin$_{4n+2}(\mC) $ & $\mu_4$   \\
$E_6(\mC)$  & $\mu_3$   \\
$E_7(\mC)$  & $\mu_2$   \\
  \hline
\end{tabular}
\\
\vspace{3mm}
\texttt{Table 4}\\
Centers of universal covering groups
\\
($\mu_N=\mZ/N\mZ$)
\end{center}
\vspace{5mm}

The factor-group $P^\vee/Q^\vee$ is isomorphic to the center $\clZ(\bG)$ of simply-connected
group $\bG$. It is a cyclic group except $\gg=D_{4l}$.
The adjoint group  $G^{ad}$ is the factor group
\beq{adj}
G^{ad}= G^\mC/\clZ(\bar G)\,.
\eq
 Let $H^\mC$ be a Cartan subgroup (Lie$(H^\mC)=\gh^\mC)$ and $N^\pm$ be the unipotent
 subgroups corresponding to the Lie algebras $\gn^{\pm}=\sum_{\be\in R^\pm}c_\be E_\be$.
 Thus $\gg^\mC=\gn^{-}+\gh^\mC+\gn^+$.
 At the group level there is the Bruhat  decomposition
 \beq{GD}
 G^\mC=N^-\cup_{w\in W}B^\mC\,,~~(\gn^{-}=Lie(N^-))\,,
 \eq
 where $B^\mC$ is the Borel subgroup
 \beq{bs}
 B^\mC=H^\mC N^+\,,~~(\gn^{+}=Lie(N^+))\,,
 \eq
 and $W$ is the Weyl group.

\paragraph{Real forms.} Let $\varsigma$ be the involutive automorphism of the group $G^\mC$,
i.e. $\varsigma^2=Id$ and $\varsigma\neq Id$. The action of $d\varsigma$ on the Lie algebra $\gg^\mC$
defines the decomposition
\beq{cde}
\gg^\mC=\gk+\gx\,,~~d\varsigma(\gk)=\gk\,,~d\varsigma(\gx)=-\gx\,.
\eq
Here $\gk$ is the invariant Lie subalgebra $[\gk,\gk]=\gk$.
Let $K$ be the corresponding  Lie groups. Then $K$ is the fixed set subgroup with
respect to  the involution $\varsigma$. The coset space $\clX=G/K$ is the symmetric space.
\cite{Be,He}. If
$g^*=\varsigma(g^{-1}$ then
\beq{rss}
\clQ=g^*g
\eq
belongs to $\clX$.

We define three involutive automorphisms and the corresponding three types of
symmetric spaces.\\
{\bf I}. Let $d\rho$ be the involutive automorphism with the fixed point subalgebra $\gc$ is the
 maximal compact subalgebra. It can be defined by the action on $x=\sum_{\al}x_\al E_\al+\sum_jx_je_j\in\gg^\mC$ (\ref{CB}) as
 \beq{rho}
 d\rho\,:\,x_\al\to-\bar x_{-\al}\,,~ x_j\to-\bar x_j\,.
 \eq
 The corresponding basis
 of $\gc$ is (\ref{cb}). The algebra $\gc$ is real and coefficients  of $x\in\gc$ in the expansion
in these basis  are real.
The corresponding subgroup of $G^\mC$ is the maximal compact subgroup $C$ and the quotient $\clX_I=G^\mC/C$ is noncompact Riemannian symmetric space. \\
{\bf II}. Let  $\gg^\mR$ be the normal real form of $\gg^\mC$. It has the same basis as $\gg^\mC$
(\ref{CB}) with real coefficients. This subalgebra is invariant under the action
 \beq{si}
    d\si(x)=\bar x\,.
    \eq
     The symmetric space $\clX_{II}=G^\mC/G^\mR$ is non-compact
 with pseudo-Riemannian metric.\\
 {\bf V}.
 The involutive automorphisms $\si$ and $\rho$ commutes.
The automorphism $d\te=d\si\circ d\rho$ acts on $\gg^\mC$ as
$$
d\te\,:\, e_j\to-e_j\,,~E_\al\to -E_{-\al}\,.
$$
The invariant subalgebra $\gk$ in (\ref{cde})
 is the complex subalgebra $\gu^\mC$
\beq{uc1}
\gk=\gu^\mC=\left\{\sum_{\al\in R^+}c_\al (E_{\al}-E_{-\al})\,,~c_\al\in\mC\right\}
\eq
and
\beq{uc2}
\gx=\left\{\sum c_je_j+\sum_{\al\in R^+}c_\al (E_{\al}+E_{-\al})\,,~c_j\,,\,c_\al\in\mC\right\}\,.
\eq
Let $U^\mC$ be the invariant subgroup of $G^\mC$ ($\te (U^\mC)=U^\mC$).
 The quotient $\clX_V= G^\mC/U^\mC$
   is the complex pseudo-Riemannian  symmetric space.
Let $\gh^\mC$ be a Cartan subalgebra $\gh^\mC\subset\gg^\mC$. It follows from   (\ref{uc2})
that $\gh^\mC\subset\gx$. Similar the Cartan subgroup $H^\mC$ of $G^\mC$ is
\beq{csv}
H^\mC\subset\clX_V\,.
\eq

In addition, consider another two types symmetric spaces.\\
{\bf III}.
Consider the defined above maximal compact subalgebra $\gc$. The involutive automorphism $d\si$ acts
on $\gc$ as ($d\si(x)=\bar x$). It follows from (\ref{ab}) that
the invariant subalgebra $gu$ has the basis (\ref{uc1}), where the coefficients $x_\al$ are real.
Let $U$ be the corresponding Lie group. Then the quotient $\clX_{III}=C/U$ is compact Riemannian symmetric space.\\
{\bf IV}. Consider action of the involutive automorphism $d\rho$ on the subalgebra $\gg^\mR$
 The invariant subalgebra  coincides with $\gu$. It is the maximal compact subalgebra of $\gg^\mR$.
 The quotient $G^\mR/U$ is a real non-compact Riemannian symmetric space $\clX_{IV}=G^\mR/U$.
 The spaces $\clX_{III}$ and $\clX_{IV}$ are Cartan dual symmetric spaces.
The considered symmetric spaces are arranged into the Table:
$$
 \begin{tabular}{|c|c|c|}
  \hline
  \hline
 $\clX$ & $G$ & $K$ \\
  \hline\hline
 $\clX_{I}$  & $G^\mC$ & $C$ \\
 \hline
 $ \clX_{II}$ &$G^\mC$ & $G^\mR$\\
 \hline
  $ \clX_{III}$ &$C$ & $U$\\
 \hline
  $ \clX_{IV}$ &$G^\mR$ & $U$\\
 \hline
  $ \clX_{V}$ &$G^\mC$ & $U^\mC$\\
 \hline
 \end{tabular}
 $$
 \begin{center}
\texttt{Table 5}\\
Symmetric spaces
 \end{center}

The concrete forms of the invariant subgroups are presented in Table 6.
$$
 \begin{tabular}{|c|c|c|c|c|}
  \hline
  \hline
  &&&&\\
  $G^\mC$ &$G^\mR$ & $C$& $U$ & $U^\mC$ \\
  &&&&\\
  \hline
  \hline
  $\SLN$& SL$(N,\mR)$ &SU$(N)$& SO$(N)$ & SO$(N,\mC)$ \\
  SO$(2N\!+\!1,\mC)$& SO$(N\!+\!1,N)$ &SO$(2N\!+\!1)$ & SO$(N\!+\!1)\times$SO$(N)$  & SO$(N\!+\!1,\mC)\times$SO$(N,\mC)$ \\
  Sp$(N,\mC)$&Sp$(N,\mR)$ &Sp$(N)$ & U$(N)$ & GL$(N,\mC)$ \\
   SO$(2N,\mC)$&SO$(N,N)$&SO$(2N)$& SO$(N)\times$SO$(N)$ &SO$(N,\mC)\times$SO$(N,\mC)$ \\
  G$^\mC_2$& G$^\mR_2$& G$_2$& SU$(2)\times$SU$(2)$ & SL$(2,\mC)\times$SL$(2,\mC)$ \\
  F$_4^\mC$ &F$_4^\mR$&  F$_4$ & Sp$(3)\times$SU$(2)$& Sp$(3,\mC)\times$SL$(2,\mC)$\\
  E$_6^\mC$& E$_6^\mR$& E$_6$& Sp$(4)$ & Sp$(4,\mC)$ \\
   E$_7^\mC$& E$_7^\mR$ & E$_7$&SU$(8)$ &SL$(8,\mC)$ \\
    E$_8^\mC$&E$_8^\mR$ & E$_8$& SO$(16)$ &SO$(16,\mC)$ \\
  \hline
 \end{tabular}
 $$
 \begin{center}
\texttt{Table 6}\\
Groups $G^\mC$, $G^\mR$, $C$, $U$, $U^\mC$
 \end{center}
These construction can be arranged in the following diagrams.
 \beq{kug}
\xymatrix{
  G^\mC  & \ar[l]^{\si}     G^\mR \\
  C \ar[u]^{\rho}
  & U\ar[l]_{\si}\ar[u]^{\rho}
                }
\eq
and
\beq{uc}
G^\mC\stackrel{\te}{\longleftarrow}U^\mC\,.
\eq
Here arrows mean embeddings as the fixed point sets of the involutive automorphisms
of the source objects to the target objects.
There are the similar relations between symmetric spaces
%
 \beq{ssr}
 \begin{array}{ccccc}
 \clX_{II}=G^\mR\setminus G^\mC &
   & \clX_V=U^\mC\setminus G^\mC & & \clX_I=C\setminus G^\mC
   \\
  \qquad\qquad \nwarrow \rho & & \nearrow\rho\qquad\qquad \si\nwarrow & & \nearrow\si\qquad\qquad
   \\
 &\clX_{III}=U\setminus C & &  \clX_{IV}=U\setminus G^\mR
 \end{array}
 \eq

\paragraph{Dimensions of algebras.}
Let $d_j$ be the order of the invariants of algebra $\gg^\mC$ and
$\rank(\gg^\mC)=l$
 \beq{dj}
d_1=2,\ldots,d_l=h\,,~~j=1,\ldots,l\,,
 \eq
where $h$ is the Coxeter number.
In terms of invariants
\beq{dcg}
\dim_{\mC}\,G^\mC=\sum_{j=1}^l(2d_j-1)=2\sum_{j=1}^ld_j-l\,.
\eq
It coincides with the real dimension of the maximal compact subgroup $C$
 \beq{dk}
\dim_{\mR}\,C=\sum_{j=1}^l(2d_j-1)\,,
 \eq
and the normal form $G^\mR$
 \beq{dgr}
\dim_{\mR}\,G^\mR=\dim_{\mR}\,C=\sum_{j=1}^l( 2 d_j-1)\,.
 \eq
  For $\clX_I=G^\mC/C$  from (\ref{dcg}) and (\ref{dk}) we have
 \beq{dgx}
\dim_{\mR}\,\clX_I=\sum_{j=1}^l  (2 d_j-1)=\dim_{\mR}\,C\,.
 \eq
 The same dimension we have for $\clX_{II}=G^\mR/C$
 \beq{dgx1}
\dim_{\mR}\,\clX_{II}=\sum_{j=1}^l  (2 d_j-1)=\dim_{\mR}\,C\,.
 \eq
The dimension of the compact form $U$ is
 \beq{du}
\dim_\mR\,U=\oh(\dim_\mR\,C-l)=\sum_{j=1}^l(d_j-1)\,.
 \eq
 Similarly,
 \beq{duc}
\dim_\mC\,U^\mC=\sum_{j=1}^l(d_j-1)\,.
 \eq
 Then
 \beq{d5}
\dim_{\mC}\,(\clX_V)=\dim_\mC\,(G^\mC)-\dim_\mC\,(U^\mC)=\sum_{j=1}^ld_j
\eq
From (\ref{dgr})  and (\ref{du}) for the symmetric spaces is $\clX_{IV}=G^\mR/U$ and $\clX_{III}=C/U$
we have
 \beq{dgxr}
\dim_{\mR}\,\clX_{III}=\dim_{\mR}\,\clX_{IV}=\sum_{j=1}^l  d_j=\oh(\dim_{\mR}\,\clX_I+l)\,.
 \eq
  Let $R^+$ be a subset of positive roots  and
  $B^\mC$ is the corresponding Borel subgroup of $G^\mC$.
  The flag variety $\Fl^\mC$ is the homogeneous space
  $\Fl^\mC(G^\mC)=G^\mC/B$.
 It has dimension
 \beq{df}
 \dim_{\mC}\,\Fl^\mC(G^\mC)=\sum_{j=1}^l(d_j-1)\,.
 \eq
From (\ref{dgx}) we have
\beq{fss}
\dim_{\mC}\,\clX_{V}=\dim_{\mC}\,\Fl^\mC+l\,.
\eq

%


\section{Cotangent bundles and coadjoint orbits}

\setcounter{equation}{0}

Let $G$ be a real simple Lie group and $\gg$ its Lie algebra.
We identify $\gg$ with the Lie coalgebra $\gg^*$ by means the invariant metric
on $\gg$.

 The cotangent bundle $T^*G\sim T\,G$ to the group $G$
  is identified with the pair
 \beq{cob}
 T^*G=\gg\times G=\{\zeta\in \gg\,,\,g\in G\}\,.
 \eq
 using its trivialization.
It is equipped with the symplectic form
  \beq{sfcb}
\om=D( \zeta,D g g^{-1})\,,~~\zeta\in\gg\,,~g\in G\,.
  \eq
The form is invariant under the left action of $G$
  \beq{ag1}
\zeta\to f\zeta f^{-1}\,,~~g\to fg
  \eq
producing the left moment
\beq{rm1}
\mu^L(\zeta,g)=\zeta\,.
\eq
The right action
\beq{ag5}
\zeta\to\zeta \,,~~g\to gf^{-1}\,,~~f\in G
\eq
leads to the moment
\beq{lm1}
\mu^R(\zeta,g)=g^{-1}\zeta g\,.
\eq
\\
\noindent
\noindent \textbf{Co-adjoint orbits of complex groups}\\
There are two ways to construct the co-adjoint $G^\mC$ orbits from the cotangent bundle  $T^*G^\mC$.

1.  Let $G=G^\mC$ be the complex simple Lie group. Assume that in (\ref{ag1}) $f\in B^\mC$ - the Borel  subgroup (\ref{bs}).
Remind that the quotient $B^\mC \setminus G^\mC$ is the flag variety $Fl^\mC$.
 The moment $\mu^L$ (\ref{lm1})
 of the left $B^\mC$-action takes value in the subalgebra
$\gb^*=\gh^\mC\oplus\gn^-$ and is equal to
\beq{mf}
\mu^L(\zeta,g)=Pr|_{\gb^*}\zeta\,.
\eq
 Fix its value as
\beq{mf1}
Pr|_{\gb^*}(\zeta )=\nu\in\gh^\mC\,.
\eq
 It means that solution of the moment equation is
  \beq{sme}
\zeta =\nu+\xi\,, ~~\xi\in\gn^+\,,
  \eq
where $\nu$ is fixed and $\xi$ is an arbitrary element of $\gn^+$.
It follows from (\ref{ag1}) and (\ref{sme}) that the symplectic quotient
 $B^\mC\setminus\!\setminus_\nu T^*G^\mC$ is defined by the pairs
 \beq{bo}
B^\mC\setminus\!\setminus_\nu T^*G^\mC=\{(g,\zeta)\,|\,1.\,g\sim fg\,,\,f\in B^\mC\,,~~2.\,\zeta~{\rm satisfy}~(\ref{sme})\}\,.
\eq

 The reduced symplectic
manifold is
  \beq{rcb}
B^\mC\setminus\!\setminus_\nu T^*G^\mC=B^\mC\setminus(\mu^L)^{-1}(\nu)\,.
  \eq
   Following (\ref{GD}) fix the gauge by taking $g=z\in N^-$.

 Notice, that the coadjoint action of $B^\mC$ on $\xi\in\gn^+$ in (\ref{sme})
 is the affine action due to the $\nu$ term. The group $B^\mC$ acts free on $\gn^+$.
It means that the symplectic quotient $B^\mC\setminus\setminus_\nu T^*G^\mC$  is the principal homogeneous space
$PH/T^*(B^\mC \setminus G^\mC/)$ over the cotangent bundle to the flag variety
 $\Fl^\mC=B^\mC \setminus G^\mC$
\beq{phs}
B^\mC\setminus\!\setminus_\nu T^*G^\mC/\sim PH/T^*(B^\mC\setminus G^\mC)\,.
\eq
The cotangent bundle $T^*\Fl^\mC$ corresponds to the choice $\nu=0$.

The symplectic form on $PH/T^*(G^\mC/B^\mC)$ is
\beq{sfd}
D(\zeta,Dg g^{-1})\,,~~(\zeta =\nu+\xi)\,.
\eq
In the coordinates $(z\in N^-,\xi)$ it takes the form
$$
D(\zeta,D z z^{-1})\,.
$$
Then $(\xi\in\gn^+,\,\log g\in\gn^-)$   are Darboux coordinates on $T^*G^\mC//B^\mC$.

The form is invariant under the right $G^\mC$-action $g\to gf^{-1}$ and
$\zeta\to \zeta $. The corresponding moment
(\ref{lm1}) is
\beq{lm2}
\mu^R=g^{-1}\zeta g=g^{-1}(\nu+\xi) g\,.
\eq

2.  Assume now that in (\ref{ag1}) $f\in G^\mC$ and
 consider the left symplectic quotient $G^\mC\setminus\setminus_\nu T^*(G^\mC)$.
with the moment taking value in the Cartan subalgebra $\gh^\mC$
\beq{or}
\mu^L(\zeta,g)=\zeta =\nu\in(\gh^{\mC}) ^*\sim\gh^{\mC}\,.
\eq
Here $\nu$ is a fixed regular element of $\gh^{\mC}$. The subgroup preserving this value is the Cartan subgroup $H^\mC$.
Thus the symplectic quotient is of the form:
\beq{or2}
G^\mC\setminus\!\setminus_\nu T^*(G^\mC)=\{(\zeta,g)\,|\, \zeta=\nu\in\gh^\mC\,,~g\sim fg\,,\,f\in H^\mC\}\,.
\eq
It is the co-adjoint orbit
\beq{orb}
G^\mC\setminus\!\setminus_\nu T^*(G^\mC)\sim\clO_\nu=H^\mC\setminus G^\mC=\{\bfS=g^{-1}\nu g\}\,.
\eq
%
After substituting $\zeta=\nu$ (\ref{or}) in (\ref{sfcb}) the form $\om$ on $T^*G^\mC$ becomes the  Kirillov-Kostant on $\clO_\nu$
form
  \beq{kkf}
\om^{KK}=D(\nu,D g g^{-1})\,.
  \eq
  The form is invariant under the right $G^\mC$ action (\ref{ag1}).
  This transformation generates the moment (\ref{lm2})
  \beq{rmo}
  \mu^R=g^{-1}\nu g
  \eq
  (compare with (\ref{lm2})).

 The representation of the orbit (\ref{orb}) is isomorphic to (\ref{phs}).
To prove it fix the gauge action (\ref{ag1})  by the choice $\xi=0$ in (\ref{sme}).
The left action of the Cartan subgroup $H^\mC$ preserves the gauge.
It means that the principal homogeneous space
$PH/T^*(G^\mC/B)$ is isomorphic to the $G^\mC$-coadjoint orbit.

Let  $\zeta=\sum S^aT_a$ be the expansion in the basis of the algebra $\gg^{\mC}$. Then
the Poisson brackets corresponding to the form $\om^{KK}$ is the Lie-Poisson brackets
\beq{lb}
\{S^a,S^b\}=C^{ab}_cS^c\,,
\eq
where $C^{ab}_c$ are structure constants of the algebra $\gg^\mC$.
These brackets are degenerated on the algebra $\gg^{\mC}$. Fixing $l$ Casimir functions
\beq{fc}
c_1(\bfS)=\nu_1\,,\ldots, c_l(\bfS)=\nu_l\,
\eq
we come to the non-degenerated brackets on the orbit $\clO_\nu$ (\ref{orb}).
The dimension of the orbit is
   \beq{dor}
 \dim\,(\clO_\nu)=2(\sum_{j=1}^ld_j-l)=\dim\,G^\mC-l\,,
     \eq
see (\ref{dcg}).
 \bigskip

\noindent \emph{\textbf{Cotangent bundles to symmetric spaces }}

Let $G$ be a  simple Lie group and $K$ its subgroup, which is invariant under the action of
an involutive automorphism $\varsigma$ ($\varsigma(f)=f$ for $f\in K$). The subgroup $K$ is reductive.
Consider  the
symmetric space $\clX=K\setminus G=\{Kg\}$. The cotangent bundle
 $T^*\clX$ can be identified with
 the symplectic quotient $K\setminus\!\setminus T^*G$. As above  the symplectic form on $T^*G$
 is $\om$ (\ref{sfcb}).
The form is invariant under the left $K$-action
\beq{ag2}
\zeta\to f\zeta f^{-1} \,,~~g\to fg\,,~~f\in K\,.
  \eq
    The symplectic
quotient
\beq{ssq}
K\setminus\!\setminus T^*G\sim T^*\clX
\eq
 is defined by the moment map
constraints
\beq{mcc}
\mu^L=\zeta |_{\gk}=0\,, ~~(\gk={\rm Lie}\,(K))\,.
\eq
Thus, $T^*\clX$ is the set of pairs $(g,\zeta)$ with the equivalence relations
\beq{pv}
(g,\zeta)\sim(fg, f\zeta f^{-1})\,,~~ f\in K\,,~~\zeta~{\rm satisfies}~ (\ref{mcc})\,.
\eq
Since $f\in K$, $f\zeta f^{-1}|_{\gk}=0$.

 The symplectic form on $T^*\clX$ coincides with (\ref{sfcb})
\beq{sfq}
\om=D(\zeta,Dg g^{-1})\,.
\eq
Consider the element
\beq{xr}
\bfX=g^{-1}\zeta g\in\gg=Lie(G)\,.
\eq
In this terms $\om$ (\ref{sfq}) assumes the form
\beq{sfq1}
\om=D(\bfX,g^{-1}Dg )\,.
\eq
The form $\om$ (\ref{sfq}) or (\ref{sfq1}) is non-degenerated on  $T^*\clX$.

 Let $g^*=\varsigma(g^{-1})$. It follows from (\ref{ag2}) that
  $T^*\clX$ is defined by the gauge-invariant variables
\beq{cav}
\clP=g^{-1}\zeta(g^*)^{-1}\,,~~\clQ=g^*g\,,
\eq
where $\zeta$ satisfies (\ref{mcc}).
Note that $\clP^*=\clP$ and $\clQ^*=\clQ$. In these variables
\beq{xr1}
\bfX=\clP\clQ\,.
\eq

For the Lie algebra $\gg$ there is the Cartan decomposition $\gg=\gk+\gx$
 (\ref{cde}), $(\gk=$Lie$(K)$ and
$\gx$ is the tangent space to $\clX=\{Kg\}$ at the point $g=Id$. It is orthogonal with respect to the invariant metric. Its restriction  on  $\gx$ is non-degenerate.
 Let $e_j$ $(j=1,\ldots,\dim\,\gp)$ be a basis in  $\gp$ with the pairing $(e_j,e_k)=\ka_{jk}$. Let $\clQ=\sum_j\clQ^je_j$, $\clP=\sum_j\clP^je_j$.
 In these variables the Poisson algebra on $T^*\clX$ is canonical
 \beq{pb2}
 \{\clP^j,\clQ^k\}=2\ka^{k}\de_{jk}\,,~~(j,k=1,\ldots,\dim\,\gp^\mC)\,.
 \eq
This relation follows from the Poisson structure on $T^*G=(\zeta,g)$:
\beq{pbt}
1.\,\{g,g\}=0\,,~~2.\,\{\zeta,g\}=g\,,~~3.\,\{\zeta^a,\zeta^b\}=C^{ab}_c\zeta^c,
\eq
where $C^{ab}_c$ are the structure constants of the Lie algebra $\gg$.

 The symplectic form  on $T^*\clX$ in terms of the  Darboux variables takes the form
\beq{caf1}
\om^X=\oh(D\clP,D\clQ)\,.
\eq
It follows from (\ref{cav}) that
\beq{ox}
\om^X=\oh(\om(\zeta,g)-\om(\zeta,(g^*)^{-1}))\,,
\eq
where $\om$ is (\ref{sfq}).

 The symplectic form (\ref{sfq1}) on $T^*\clX$
 is invariant under the right action of the group $G^\mC$
(\ref{ag5}),
\beq{ras}
g\to gf^{-1}\,,~~\bfX\to f\bfX f^{-1}\,.
\eq
 Similarly to (\ref{rm1}) the moment corresponding to this action is
\beq{rm0}
\mu^R(\zeta,g)=g^{-1}\zeta g\in\gg^\mC\,,~~{\rm where}~\zeta|_{\gk}=0
\eq
(see (\ref{mcc})),  or
  \beq{mla1}
  \mu^R\stackrel{(\ref{xr})}=\bfX\stackrel{(\ref{xr1})}=\clP\clQ\,.
  \eq
This construction is applicable to the symmetric spaces from Table 1.

Consider the special case $T^*\clX_V=U^\mC\setminus G^\mC$.
The form $\om$ (\ref{sfq}) is the result of the symplectic quotient of the
cotangent bundle under the left action of the subgroup $U^\mC$. Since $H^\mC\notin U^\mC$
(\ref{csv}) one can consider the additional left action of Cartan subgroup $H^\mC$
\beq{csa}
g\to fg\,,~~\zeta\to f\zeta f^{-1}\,, ~~f\in H^\mC\,.
\eq
The symplectic reduction is defined by the moment constraint equation
\beq{pao}
\mu^L=Pr\zeta \,|_{\gh^\mC}=\nu
\eq
and the gauge invariant variable $\bfX$ (\ref{xr}). Therefore $\bfX=g^{-1}\nu g\in\gg^\mC$
is an element of the coadjoint orbit $\clO_\nu$. It means that
\beq{oga}
\clO_\nu=\{\bfX=g^{-1}\nu g\,|\,g\in G^\mC\}=H^\mC\setminus\setminus_\nu T^*\clX_V\,.
\eq
\bigskip


\end{document}


\end{document}